\documentclass[review]{elsarticle}
\usepackage[margin=2.0cm]{geometry}
\usepackage{lineno,hyperref}
\usepackage{subfig}
\usepackage{longtable}
\usepackage{caption}
\modulolinenumbers[5]
\usepackage{algorithmic}
\usepackage{longtable}
\usepackage{multicol}
\usepackage{multirow}
\usepackage{graphicx}
\usepackage{xcolor}
\usepackage{lscape}

\usepackage{fancyhdr}
\journal{Journal of \LaTeX\ Templates}

\makeatletter \def\ps@pprintTitle{  \let\@oddhead\@empty  \let\@evenhead\@empty  \def\@oddfoot{\hfill\thepage}  \def\@evenfoot{\thepage\hfill}} \makeatother

\begin{document}
\begin{frontmatter}

\title{ Machine Learning-based Lung and Colon Cancer Detection using Deep Feature Extraction and Ensemble Learning}


\author[1]{Md. Alamin Talukder}
\ead{mdalamintalukdercsejnu@gmail.com}
\author[1]{Md. Manowarul Islam}
\ead{manowar@cse.jnu.ac.bd}
\author[1]{Md Ashraf Uddin}
\ead{ashraf@cse.jnu.ac.bd}
\author[1]{Arnisha Akhter}
\ead{arnisha@cse.jnu.ac.bd}
\address[1]{Department of Computer Science and Engineering, Jagannath University, Dhaka, Bangladesh}

\author[2]{Khondokar Fida Hasan}
\ead{fida.hasan@rmit.edu.au}

\address[2]{Centre for Cyber Security Research \& Innovation, RMIT University, 124 La Trobe Street, Melbourne, 3000, VIC, Australia}

\author[4]{Mohammad Ali Moni}
\ead{m.moni@uq.edu.au}

\address[4]{Artificial Intelligence \& Data Science, School of Health and Rehabilitation Sciences, Faculty of Health and Behavioural Sciences, The University of Queensland St Lucia, QLD 4072, Australia.}

\cortext[mycorrespondingauthor]{Mohammad Ali Moni}
\cortext[mycorrespondingauthor]{Md Manowarul Islam}

\begin{abstract}
Cancer is a fatal disease caused by a combination of genetic diseases and a variety of biochemical abnormalities. Lung and colon cancer have emerged as two of the leading causes of death and disability in humans. The histopathological detection of such malignancies is usually the most important component in determining the best course of action. Early detection of the ailment on either front considerably decreases the likelihood of mortality. Machine learning and deep learning techniques can be utilized to speed up such cancer detection, allowing researchers to study a large number of patients in a much shorter amount of time and at a lower cost.  In this research work, we introduced a hybrid ensemble feature extraction model to efficiently identify lung and colon cancer. It integrates deep feature extraction and ensemble learning with high-performance filtering for cancer image datasets. The model is evaluated on histopathological (LC25000) lung and colon datasets.  According to the study findings, our hybrid model can detect lung, colon, and (lung and colon) cancer with accuracy rates of 99.05\%, 100\%, and 99.30\%, respectively.  The study's findings show that our proposed strategy outperforms existing models significantly. Thus, these models could be applicable in clinics to support the doctor in the diagnosis of cancers.

\end{abstract}


\end{frontmatter}


\section{Introduction}
\label{intro}

Cancer is a broad  classification that encompasses a variety of illnesses that can damage organs in the human body. It is also a rapid reaction of aberrant cells that expand beyond their acceptable boundaries, allowing them to infect neighboring areas and travel to neighboring tissues which are known as metastasis.  Metastasis is the prime  factors of cancer-related death \citep{whonews}. The disease's prevalence has risen significantly over time, most likely as a result of increased sensitivity to particular cancers as age advances. Colorectal cancer, which affects the colon and rectum, is one of the major induces of death. Lung cancer is probably the most commonly recognized malignancy (12.2\%), but colorectal cancer, which affects the colon and rectum, is also the leading trigger of death (10.7\%) \citep{cancertoday}. It is one of the major drivers of fatality globally, although its impact is not  evenly distributed \citep{vineis2014global}. When it comes to identifying cancer symptoms, there are quite a variety of approaches. With the advancements, the amount of data stored in archives is growing by the day \citep{godkhindi2017automated}. It is extremely challenging to analyze a huge quantity of data and extrapolate using traditional approaches \citep{yildirim2022classification}.
The increased availability of healthcare data provides scientists with a fresh opportunity to improve existing approaches for further comprehensive clinical analysis \citep{sarwinda2020analysis}. In the medical profession, machine learning and deep learning are commonly utilised technologies for analyzing biomedical data \citep{park2018machine} \citep{das2020experimental} \citep{baldi2018deep} \citep{lenz2019unsupervised}.

Machine learning (ML) is a limb of AI technologies that arose from the field of pattern identification and cognitive acquisition concepts. It produces mechanisms that effectively adapt through a vast group of information and generate forecasts based on historical evidence \citep{park2018machine}. In several sectors where developing explicit techniques with acceptable efficiency seemed hard and nearly impossible, machine learning had indeed been successfully employed with remarkable outcomes \citep{hussain2021machine} \citep{assegie2021optimized} \citep{xue2021identifying}. 
Deep learning (DL), on the contrary, is a sophisticated based pattern classification technology that has shown exceptional results in feature extraction, object detection and voice recognition, and other domains that require multilevel data processing \citep{pyrkov2018extracting}. It can extract meaningful significant patterns from images and has been demonstrated to obtain state efficiency, occasionally outperforming humans \citep{ciregan2012multi}. 
Transfer learning (TL) entails repurposing previously acquired knowledge to tackle a specific situation \citep{kandaswamy2016high}. It can be used in two ways: as a baseline algorithm, which is used to train the image dataset and determine performance \citep{hijab2019breast} \citep{chougrad2020multi} \citep{khan2019novel}; as a feature extractor, which is used to extract features from image datasets and then use machine or deep learning algorithms to determine performance \citep{deniz2018transfer} \citep{de2019double} \citep{abbasi2020detecting} \citep{phankokkruad2021ensemble}.
Ensemble learning relies upon the combination of many learning models that are purposefully developed as well as integrated to solve issues like categorization \citep{polikar2012ensemble}. It is a kind of machine learning that aims to upgrade predicted efficiency by combining several models. Moreover, it's a significant subject for increasing base classification models \citep{onan2015performance}.

A number of papers have been conducted in the literature, each of which is predicated on a distinct strategy for detecting cancer \citep{hatuwal2020lung}  \citep{tasnim2021deep}  \citep{liang2020identification} \citep{chen2021detection}. A wide variety of medical images have already been identified and signified with the help of machine learning procedures. DL strategies have made it possible for machines to assess exalted dimensional information such as pictures, multimodal pathology scans, and video files \citep{adeoye2022comparison} \citep{alsinglawi2022explainable} \citep{tsai2022machine}. Numerous supervised machine learning techniques have been produced, as they are extremely adept at handling biological images \citep{bukhari2020histological} \citep{wu2017small} \citep{papp2021supervised} \citep{urbanos2021supervised}.

To classify lung and colon cancer efficiently, we need to train our model with a large dataset where all cancer possibilities are included. To do so, our research emphasizes designing a resourceful and well-organized framework where we can take large histopathological image (HPI) datasets (LC25000) and a number of pre-processing methods are applied to datasets. k-fold cross-validation, and feature extraction, all are applied to it. To build a robust model, we have used an ensemble method at the end of our implementations getting better performance in cancer classifications.

The performance of our proposed hybrid model will be evaluated using high-performance filtering (HPF) with  combination of ensemble learning. We have evaluated a variety of performance indicators in our investigation including accuracy, recall, precision , f1-score, MAE, MSE, RMSE, confusion matrix, AUC score and the ROC Curve. The classifier algorithms in the proposed framework can detect cancer (lung and colon) having an accuracy rate of above 99\%. 

The following are the major accomplishments of this study.
\begin{itemize}
    \item First of all, to detect lung and colon cancer, we proposed an efficient hybrid technique that combines deep feature extraction, high-performance filtering, and an ensemble learning strategy.
    \item Secondly, A deep feature extraction technique is proposed that is uniquely applied using different transfer learning models to extract features. 
    \item Finally, we analyze our proposed model on HPI datasets using a set of metrics and showed how to diagnose lung and colon cancers, thereby contributing to the new knowledge of AI-based applications.
   
\end{itemize}

The remainder of the paper is arranged as follows: Section \ref{sec:Related} leads a literature review to identify the state-of-art of lung and colon cancer diagnoses using machine learning. Section \ref{sec:Method} gives the methodology and the dataset description of the study.  The experimental setup and performance evaluation are presented in Section \ref{sec:experimental} and Section \ref{sec:evaluation}, respectively before concluding the paper in section \ref{sec:conclusion}.

\section{Literature Review}
\label{sec:Related}
Categorizing histopathological image datasets of many cancer kinds, such as breast, lung, colon, colorectal, and skin cancer, has received much attention. Various approaches are utilized using ML, DL, and TL to detect lung and colon cancer. 
 
\subsection{Lung and colon cancer using computer-aided diagnosis (CAD) approach}
A CAD method was introduced by \citep{nishio2021homology} to categorize histopathology images of lung tissues automatically. They evaluated the effectiveness of eight ML algorithms on two datasets using traditional texture analysis (TA) and homology-based image processing (HI) to visual feature extraction. The CAD system with HI outperformed the TA system in both datasets. As a result, they determined that HI was much more beneficial for CAD systems than TA and that a precise CAD system for lung tissues might be developed. Similarly, \citep{mangal2020convolution} created a CAD system by examining digital pathology images for recognizing lung and colon cancer utilizing a convolutional neural network (CNN). In comparison to classical ML models as well as deep CNN models utilizing TL trained on a similar collection though utilizing state-of-the-art feature descriptors, their empirical findings on the LC25000 indicated better accuracy of 97.89\% for lung and 96.61\% for colon gained by CNN. Moreover, \citep{shandilya2022analysis} devised a CAD technique for categorizing histological pictures of lung tissues. They utilized a publicly accessible dataset of histological photos of lung tissue for the design and validation of CAD. Multi-scale processing was performed to extract image features. Furthermore, a comparative analysis was conducted using seven pre-trained hyper-tuned CNN models for lung cancer prediction where, ResNet101 provided the highest accuracy of all, at 98.67 percent. This study will assist researchers in developing more effective CNN-based lung cancer detection models. 

\subsection{Lung and colon cancer using  CNN approach}
A CNN approach was proposed by \citep{hatuwal2020lung} for detecting lung cancer using histopathology pictures. Their model revealed a considerable improvement in classification accuracy and got training and validation accuracy of 96.11\% and 97.20\%, respectively. Similarly, \citep{tasnim2021deep} introduced three CNN models to analyze colon cell imaging data. The models were trained and tested at various epochs to estimate the learning rate. The average pooling and max-pooling layers are revealed to be 95.48\% and 97.49\% accurate, respectively. The MobileNetV2 exceeds the other two models with the highest accuracy of 99.67\% and the lowest loss rate of 1.24\%. It would be developed in collaboration with medical professionals at hospitals or clinics that specialize in colon cancer treatment. \cite{qasim2020convolutional} produced a CNN model for predicting colon cancer distinguished by speed and accuracy with a small number of parameters. Their model was based on two routes, each of which was in charge of building 256 feature maps to enhance the number of features at various levels and so raise the accuracy and sensitivity. The visual geometry group (VGG16) model was created and trained within the same dataset to assess the effectiveness of the proposed scheme. The acquired accuracy is 99.6\% for the proposed model and 96.2\% for VGG16. The performance indicated that it was useful in identifying colon cancer. On the other hand, \citep{sikdersupervised} proposed a unique method for segmenting, recognizing, classifying, and detecting various types of malignant cells in both MRI and RGB scans. They used a CNN model and a SegNet approach with a morphological operation that was superior to the typical SegNet model to minimize training time and boost segmentation results. The proposed technique produced an average accuracy rate of 93 percent in detecting cancer cells from multiple cancer datasets. They were able to overcome the limitations of employing separate cancer identification techniques for MRI and Histopathology data. 

\subsection{Lung and colon cancer using DL and ML approach}
An ML approach to diagnosing lung and colon cancer based on DL was proposed by \citep{masud2021machine} that analyzed pathological photos of lung and colon cancers to identify five different types of tissues. They extracted the features using 2D Fourier and 2D Wavelet (2D FW) feature extraction process and combine the features and trained them with a CNN model. The obtained findings revealed that the presented architecture has the highest accuracy of 96.33 cents in identifying cancer tissues. The use of this approach will aid healthcare practitioners in the development of a computerized and efficient method for detecting different forms of lung and colon cancers. Similarly, \citep{sarwinda2020analysis} evaluated the effectiveness of deep feature extraction for colorectal cancer diagnosis using two TL models as well as traditional ML algorithms. The findings revealed that employing DenseNet121 to extract the features outperforms ResNet50 for all ML algorithms and achieved 98.53\% accuracy utilizing the KNN algorithm. The developed approach proved adequate for identifying colon tissue. To enhance picture classification efficiency, \citep{bansal2021transfer} presented an approach integrating deep features generated with VGG19 with other customized feature extraction techniques, such as SIFT, SURF, ORB, and the Shi-Tomasi corner detector algorithm. Additionally, the collected features from these methods were categorized using a variety of ML algorithms. The empirical findings showed that Random Forest (RF) with the cooperative features beat other classifiers with 93.73 percent accuracy. According to the findings, a single feature extractor is insufficient to provide consistent outcomes. As a result, for picture classification, a mixed method employing DL features as well as classical handmade features is preferable. \cite{tougaccar2021disease} introduced a DL model where the image classes were built from the ground up using the DarkNet19 model. The feeble features were selected from the feature set retrieved from the DarkNet19 model using the Equilibrium and Manta Ray Foraging optimization approach and separated from the rest of the set's features, yielding an optimal feature set. The Support Vector Machine (SVM) approach was accustomed to aggregating and categorizing the suitable characteristics generated by the two optimization algorithms utilized. The overall classifier performance percentage was 99.69\%. It was discovered that combining the complementary approach with certain optimizing strategies increased the dataset's categorization ability. To determine lung and colon cancer using histopathological images with improved augmentation strategies, \citep{garg2020prediction} proposed a method by modifying 8 TL models. According to the findings, the 8 models achieved notable accuracy ranging from 96 percent to 100 percent accuracy. To visualize the focus images of CNN models, SmoothGrad and GradCAM were then employed that had previously been learned to differentiate between cancerous and benign images. 

\subsection{Lung and colon cancer using Hybrid approach}
In \citep{phankokkruad2021ensemble}, for diagnosing lung cancer three TL models were utilized. The findings exhibited that the proposed models attained accuracy levels of 62 percent, 90 percent, and 89 percent for VGG16, ResNet50V2, and DenseNet201, respectively. Lastly, a combination of the three proposed CNN models called the ensemble model exceeded the other models with a validation accuracy of 91\%. The employment of the ensembles of the TL model, which enhances efficiency, is beneficial in overcoming the challenge of diagnosing lung cancer. Another, \citep{chen2021detection} proposed a hybrid classifying framework that combines the Inception v3, hog, and daisy extracting features modules to distinguish lung cancer and normal tissue from lung pathology images. The experimental findings showed that the proposed model achieved an accuracy of 99.60 percent, which was a gratifying outcome greater than other comparable trials. It demonstrated that hybrid deep learning models may be used to evaluate a reliable cancer diagnosis approach. On the other hand, \citep{liang2020identification} developed a shearlet transform-based multi-scale feature fusion (MFF-CNN) model to recognize histopathology images of colon cancer. It achieved a 96 percent detection performance for colorectal after feature learning and feature fusion. Its improvement offers innovative opportunities for cancer screening instantaneously, with objectivity as well as accuracy. \cite{yildirim2022classification} created a new method for identifying colon cancer images called MA ColonNET. The algorithm reached a 99.75 cent accuracy rate for detecting and classifying colon cancer which will help to avoid human errors that are common in older methods. They demonstrated that the presented methodology can identify colon cancer faster than originally thought. The therapy strategy will be more effective as a result of this as well. Furthermore, \citep{adu2021dhs} offered the DHS-CapsNet, a revolutionary dual horizontal squash capsule network for classifying lung and colon tumors on histopathological pictures. Encoder feature fusion (EFF) and a unique horizontal squash (HSquash) algorithm were used to build DHS-CapsNet. The collected feature from the 2-lane convolutional layers was aggregated by the EFF, which provided valuable insights for greater accuracy. The claiming prediction error for the DHS-CapsNet was 0.77 percent, as opposed to 14.45 percent for the regular CapsNet. It enhanced CapsNet which can be used as a CAD approach to aid clinicians in diagnosing lung and colon cancer. The related paper summary is shown in Table \ref{tab:related_summary}.

\begin{table}[]
\centering
\begin{tabular}{lllll}
\hline
SL. No & Author & Dataset & Approach & Accuracy(In \%) \\ \hline
1 & \cite{nishio2021homology} & Histopathology & CAD & 99.23 \\
2 & \cite{mangal2020convolution} & Histopathology & CAD & \begin{tabular}[c]{@{}l@{}}97 (Lung)\\ 96 (Colon)\end{tabular} \\
3 & \cite{shandilya2022analysis} & Histopathology & CAD & 98.67 \\
4 & \cite{hatuwal2020lung} & Histopathology & CNN & 97.2 \\
5 & \cite{tasnim2021deep} & Histopathology & CNN & 99.67 \\
6 & \cite{qasim2020convolutional} & Histopathology & CNN & 99.6 \\
7 & {\citep{sikdersupervised}} & \begin{tabular}[c]{@{}l@{}}Brain MRI Image\\ BreCaHAD \\ SN-AM\\ Histopathology\end{tabular} & CNN & \begin{tabular}[c]{@{}l@{}}98.59 (IoU)\\ 91.16 (IoU)\\ 93.02 (IoU)\\ 90.24 (IoU)\end{tabular} \\
8 & \cite{masud2021machine} & Histopathology & 2D FW+ CNN & 96.33 \\
9 & \cite{sarwinda2020analysis} & Histopathology & DenseNet121 + KNN & 98.53 \\
10 & \cite{bansal2021transfer} & Caltech101 & VGG19 + RF & 93.73 \\
11 & \cite{tougaccar2021disease} & Histopathology & DarkNet19 + SVM & 99.69 \\
12 & \cite{garg2020prediction} & Histopathology & \begin{tabular}[c]{@{}l@{}}VGG16\\ ResNet50\\ InceptionV3\\ InceptionResNetV2\\ MobileNet\\ Xception\\ NASNetMobile\\ DenseNet169\end{tabular} & \begin{tabular}[c]{@{}l@{}}98 \\ 96\\ 100\\ 100 \\ 100\\ 100\\ 97\\ 100\end{tabular} \\
13 & \cite{phankokkruad2021ensemble} & Histopathology & Ensemble Model & 91 \\
14 & \cite{chen2021detection} & Histopathology & Ensemble Model & 99.6 \\
15 & \cite{liang2020identification} & Histopathology & MFF-CNN & 96 \\
16 & \cite{yildirim2022classification} & Histopathology & MA ColonNET & 99.75 \\
17 & \cite{adu2021dhs} & Histopathology & DHS-CapsNet & 99.23 \\ \hline
\end{tabular}%

\caption{Related work summary}
\label{tab:related_summary}
\end{table}
 
\section{Methodology}
\label{sec:Method}
In this section, we have covered image pre-processing, deep feature extraction, high-performance filtering, and ensemble learning method as well as our proposed model. We also give a basic rundown of the ML techniques that are used to detect lung and colon cancer. 

\begin{figure*}[!htbp]
	\centering
	{\includegraphics[scale=.20]{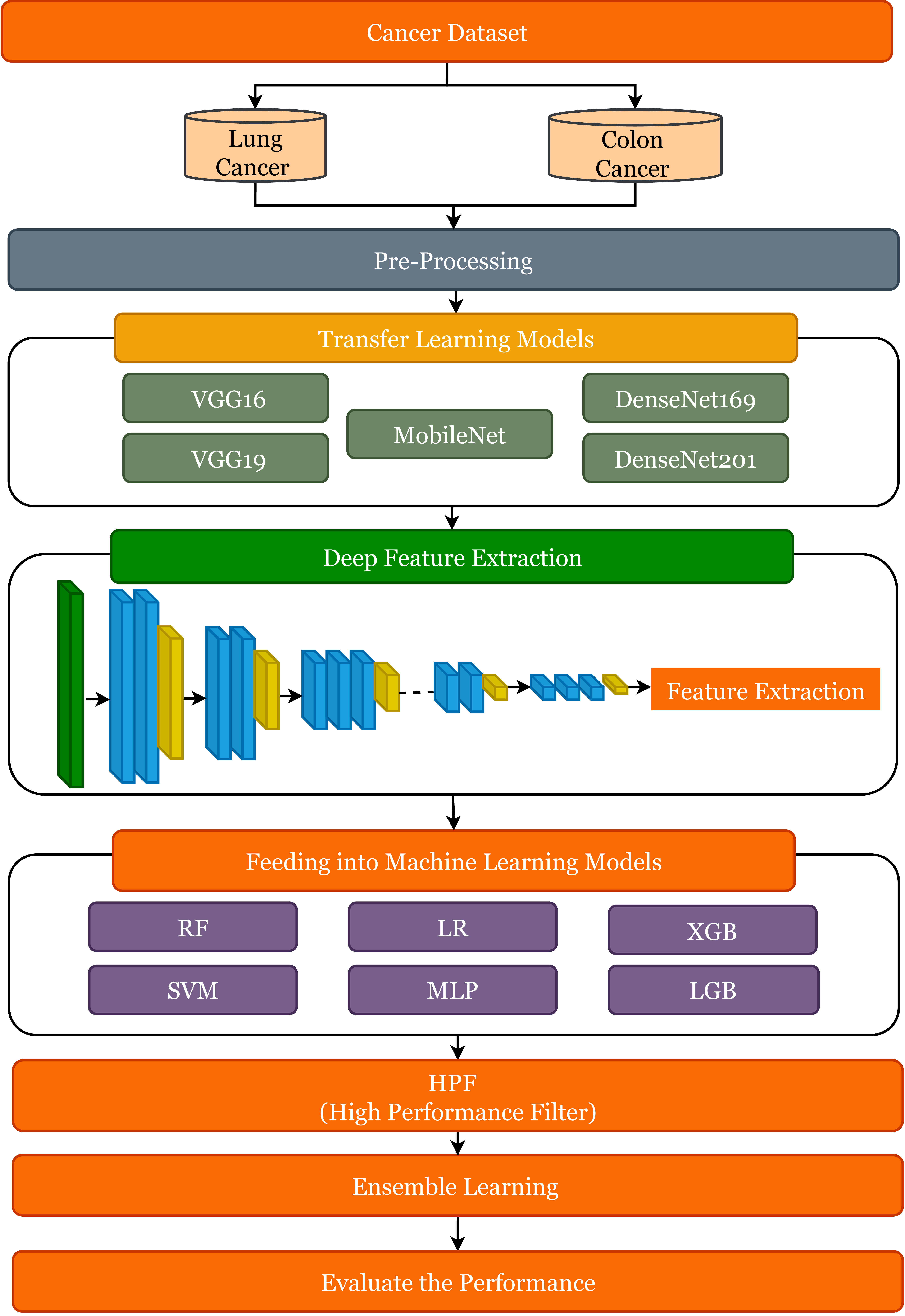}}
\caption{The proposed hybrid model for lung and colon cancer detection}
\label{fig:proposal}
\end{figure*}

We developed our proposed model using feature extraction and machine learning algorithms on LC25000 lung and colon histopathological image datasets to assure a prognosis of lung and colon cancer. Figure \ref{fig:proposal} represents the schematic block diagram of our proposed paradigm. The stages of our proposed model are as follows:

\begin{itemize}
    
    \item {Step-1: }Initially, we take the histopathological image datasets (LC25000) for lung and colon and combine both (lung and colon) cancer to conduct our experiment. There are three types of lung cancers, and two types of colon cancers are available in our datasets.
    \item {Step-2: } In the pre-processing step, the pre-processing is accomplished by resizing the image into 128x128, converting the image into bgr2rgb, and then the image to NumPy array. After that, we perform feature scaling where we use the generalization method in the image; we divide easy image array values by 255, which is the maximum value (max. intensity value of an image), and perform labeling by assigning a label (0,1,2) for each image.
    \item {Step-3: } In the deep feature extraction step, we extract features from the image datasets using five different transfer learning such as VGG16, VGG19, DenseNet169, DenseNet201 and collect the features for next the step.
    \item {Step-4: } In this step, successfully extracted features are fit into the six well-known ML algorithms such as Random Forest (RF), Support Vector Machine (SVM), Logistic Regression (LR), Multi-Layer Perceptron (MLP), Extreme Gradient Boosting (XGB), and Light Gradient Boosting (LGB) to evaluate the performance.
    \item {Step-5: } In this High-Performance Filtering (HPF) step, we filter our ML algorithms based on the accuracy. By using this filtering, we select the top three algorithms for the next step. 
    \item {Step-6: } In this Ensemble Learning step, after getting the top three algorithms from HPF, we apply majority (hard) and weighted average (soft) voting classifiers on these algorithms for all TL and select the best ensemble voting classifier for each TL model. 
    \item {Step-7: } Finally, the performance is assessed based on the accuracy  for each TL and selecting the best TL model. After that, the performance metrics such as accuracy, recall, precision, f1-score, MAE, MSE, RMSE, confusion matrix, AUC score, and ROC Curve are used to evaluate the selected model of our experiment and also showed the comparison analysis with other existing models. 
\end{itemize}

\subsection{Dataset Collection}
In our proposed work, we have employed the lung and colon cancer dataset called histopathological images (LC25000) dataset. The dataset consists of two cancerous datasets called lung and colon cancer datasets.  It contains 25000 images where 10000 for colon cancer and 15000 for lung cancer images. The lung cancer dataset contains 3 labels of cells such as adenocarcinomas, squamous cell carcinoma and benign tissue. The colon cancer dataset contains 2 labels of cells, such as adenocarcinomas and benign tissue. The LC25000 dataset was created using a snippet of HIPAA-compliant as well as confirmed references, including 750 lung tissue where 250 adenocarcinomas, 250 squamous cell carcinomas and 250 benign tissue as well as 500 colon tissue where 250 adenocarcinomas and 250 benign tissue which were then augmented to create 25,000 images \citep{borkowski2019lung}.

In our experiments, we have used lung, colon, and both (lung and colon) cancer images. So, we took 10000 images consisting of 5000 images for colon tissue and 5000 images for lung tissue for both lung and colon; 2800 images for colon tissue for colon; 4200 images for lung tissue for lung. The distribution of Cancerous datasets images is shown in Table\ref{table:lc_distitubution}, Table \ref{table:cc_distitubution}, and Table\ref{table:c_distitubution}, for our research.

\begin{table}[]
\centering
\begin{tabular}{ll}
\hline
Lung Cancer                    & No. of Images \\ \hline
Benign lung   tissue           & 1400          \\ 
Lung   adenocarcinomas         & 1400          \\ 
Lung squamous   cell carcinoma & 1400          \\ 
Total                          & 4200          \\ \hline
\end{tabular}
\caption{Distribution of Lung Cancer dataset}
\label{table:lc_distitubution}
\end{table}

\begin{table}[]
\centering
\begin{tabular}{ll}
\hline
Colon Cancer            & No. of Images \\ \hline
Benign colon   tissue   & 1400          \\ 
Colon   adenocarcinomas & 1400          \\ 
Total                   & 2800          \\ \hline
\end{tabular}
\caption{Distribution of Colon Cancer dataset}
\label{table:cc_distitubution}
\end{table}

\begin{table}[]
\centering
\begin{tabular}{ll}
\hline
Cancer                    & No. of Images \\ \hline
Benign lung   tissue           & 2000          \\ 
Lung   adenocarcinomas         & 2000          \\ 
Lung squamous   cell carcinoma & 2000          \\ 
Benign colon   tissue          & 2000          \\ 
Colon   adenocarcinomas        & 2000          \\ 
Total                          & 10000          \\ \hline
\end{tabular}
\caption{Distribution of Cancer (lung and colon) dataset}
\label{table:c_distitubution}
\end{table}

\subsection{Pre-processing}
\begin{figure*}[!htbp]
	\centering
	{\includegraphics[scale=.20]{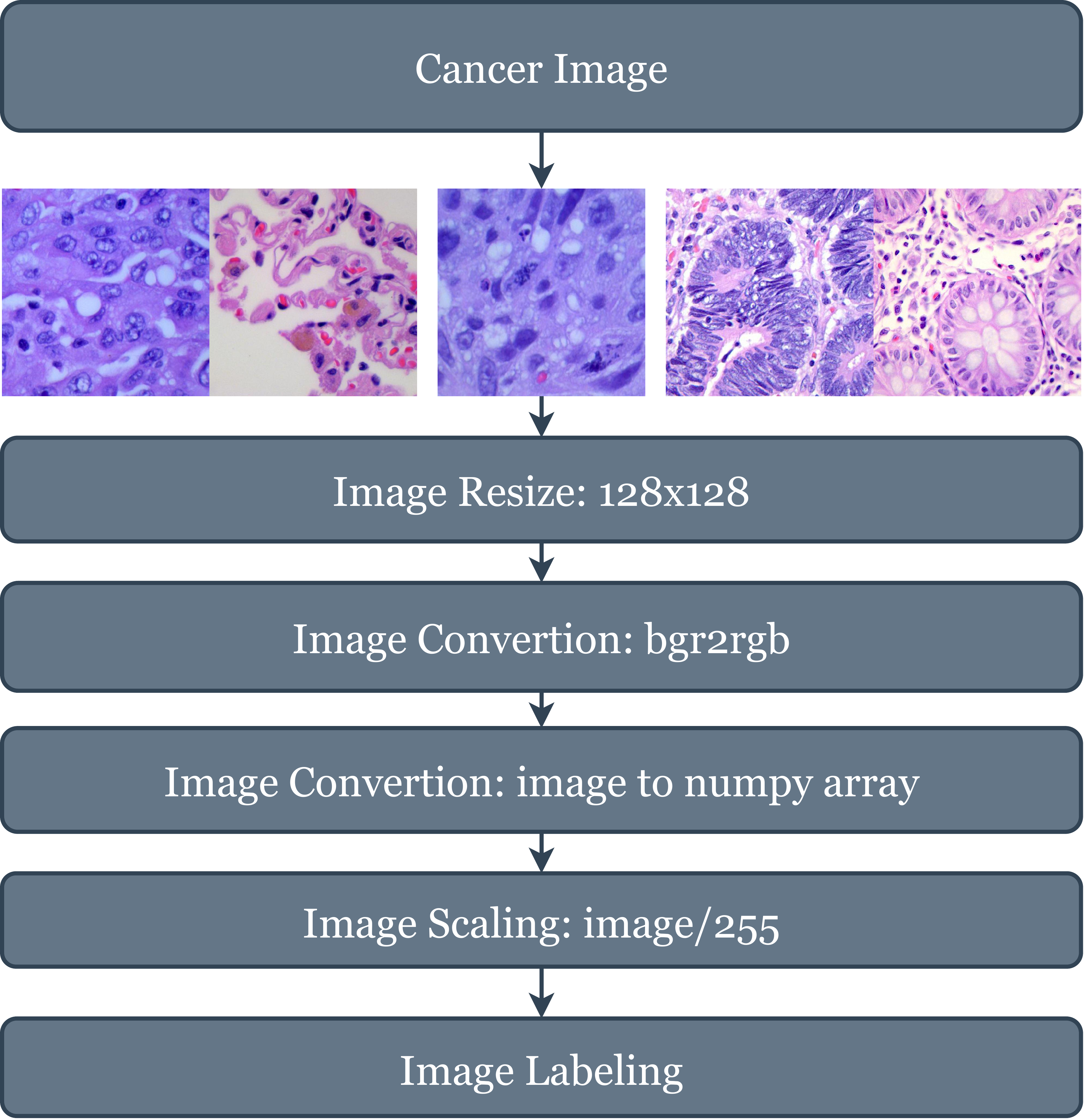}}
	\caption{Pre-processing steps}
	\label{fig:pre_process}
\end{figure*}

Initially, in our methodology, we downsized the input image to 128x128 while keeping the same image size to train the model. We transformed the scaled image into bgr2rgb and then into a NumPy array to work with picture intensity values. Then, perform scaling, a procedure that normalizes the value of image intensity values within a range (0 to 1). To reduce computational complexity, the image is scaled by dividing the image array by 255. Finally, we added the image label as Image labeling, which is an important step because it allows us to identify malignant photos. Fig. \ref{fig:pre_process} depicts the pre-processing steps of our proposed model. Table \ref{tab:label_lung} and Table \ref{tab:label_colon} show labeling for lung and colon cancer dataset, respectively.

\begin{table}[]
\centering
\begin{tabular}{lll}
\hline
Lung Cancer                    & Dataset   & Label \\ \hline
Lung adenocarcinoma            & lung\_aca & 0     \\ 
Lung benign tissue             & lung\_n   & 1     \\ 
Lung squamous   cell carcinoma & lung\_scc & 2     \\ \hline
\end{tabular}
\caption{Labeling for lung cancer dataset}
\label{tab:label_lung}
\end{table}

\begin{table}[]
\centering
\begin{tabular}{lll}
\hline
Colon Cancer         & Dataset    & Label \\ \hline
Colon adenocarcinoma & colon\_aca & 0     \\ 
Colon benign tissue  & colon\_n   & 1     \\ \hline
\end{tabular}
\caption{Labeling for colon cancer dataset}
\label{tab:label_colon}
\end{table}

\begin{figure*}[!htbp]
	\centering
	\subfloat[before]{\includegraphics[scale=.450]{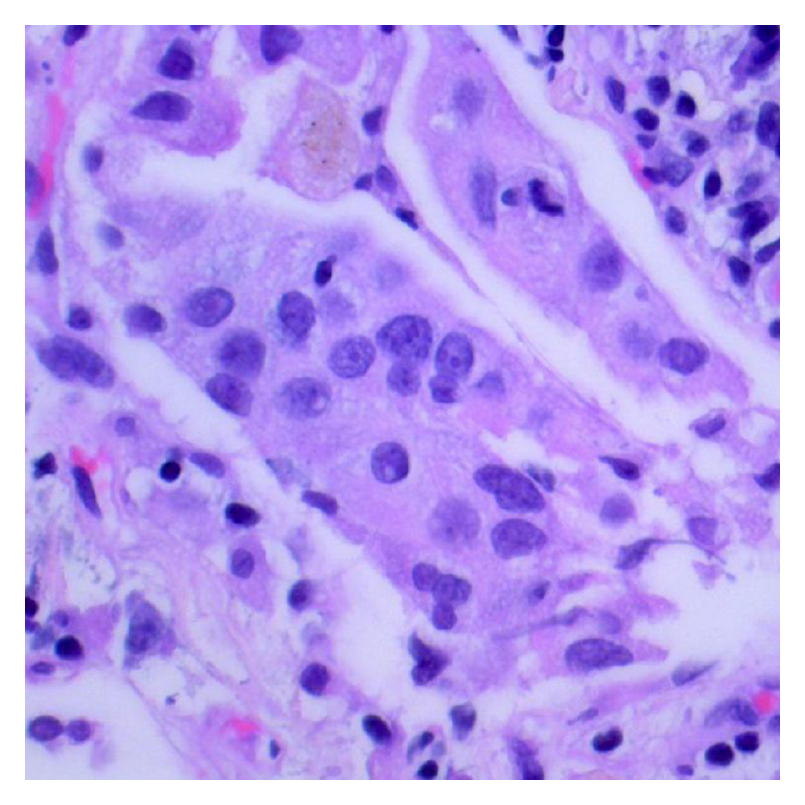}}
	\subfloat[before]{\includegraphics[scale=.450]{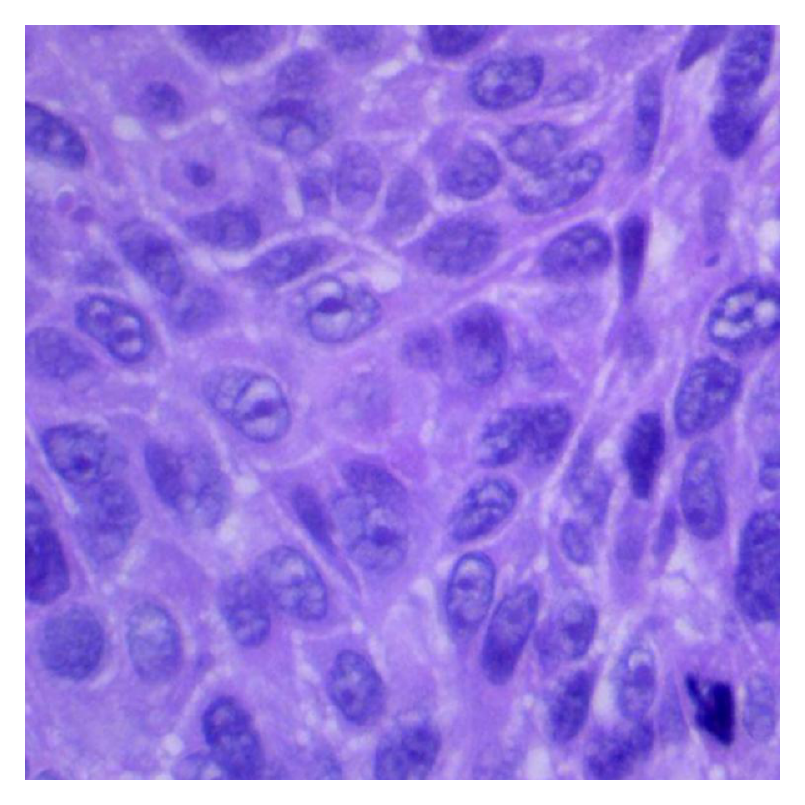}}
    \subfloat[before]{\includegraphics[scale=.450]{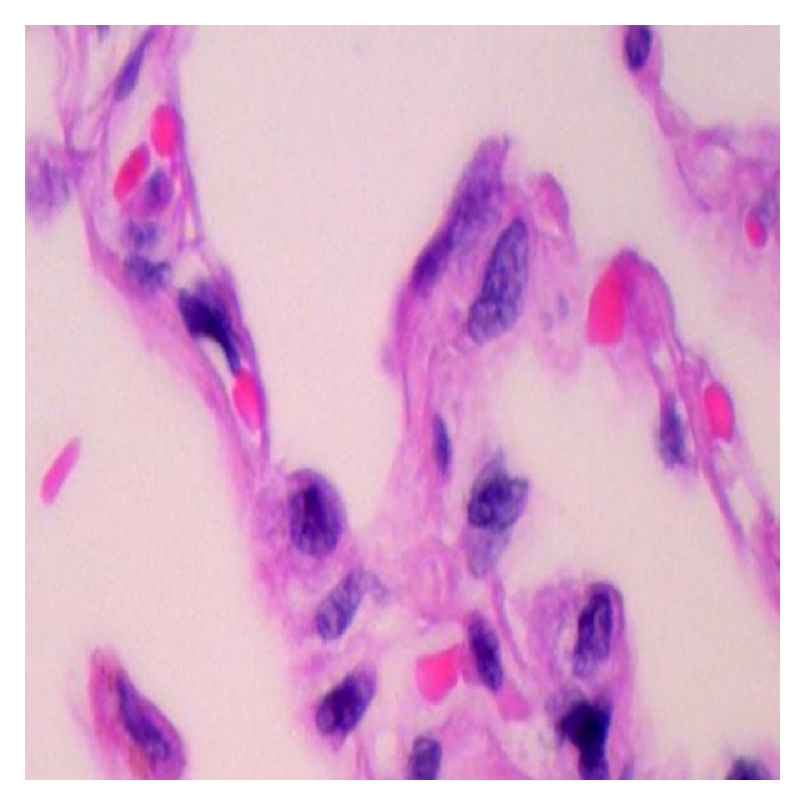}} \hspace{0.1cm}
    
	\subfloat[after]{\includegraphics[scale=.450]{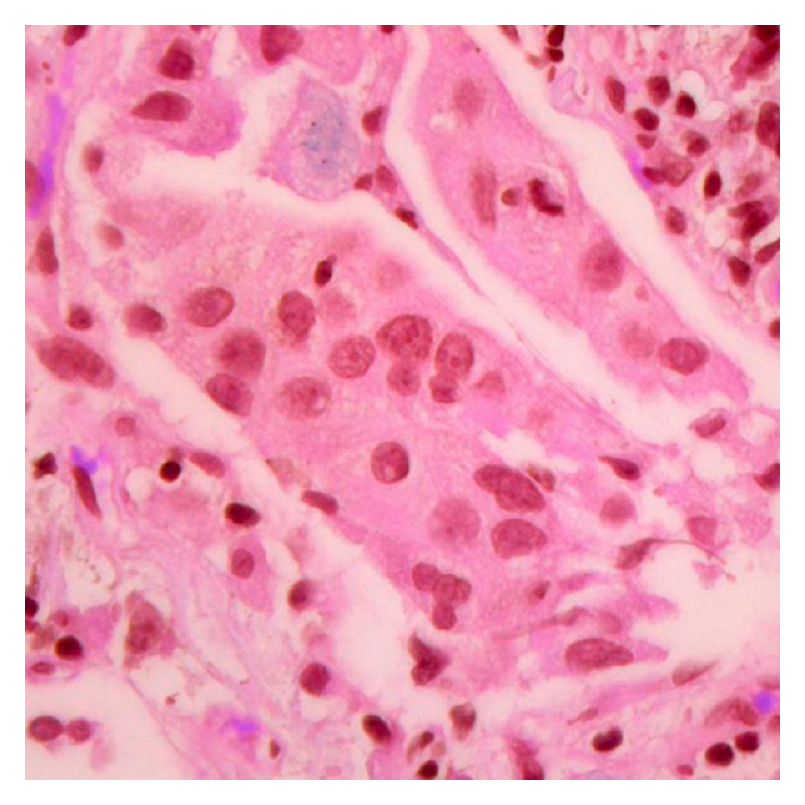}}
    \subfloat[after]{\includegraphics[scale=.450]{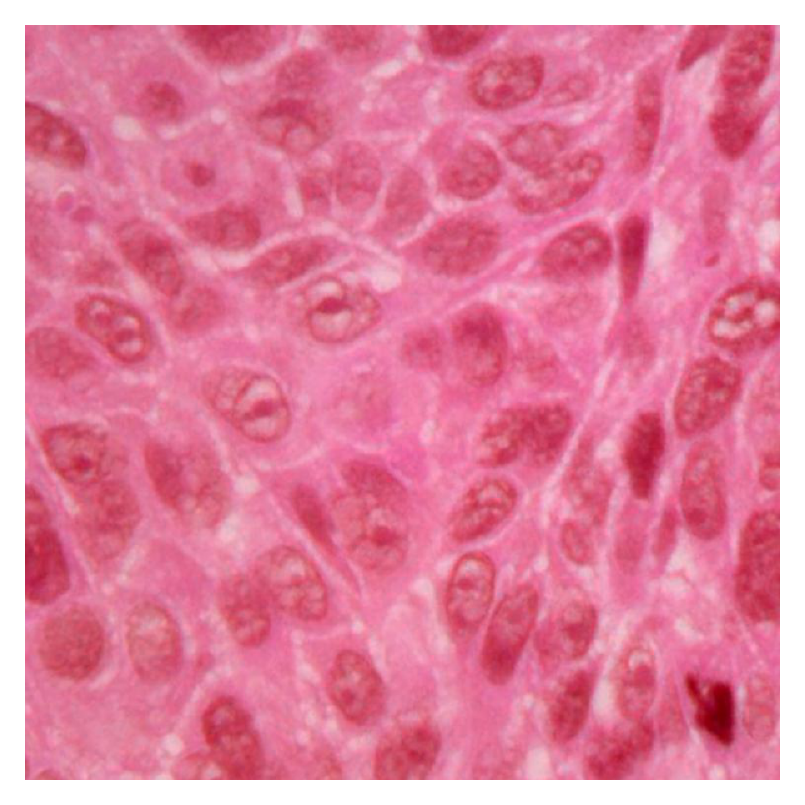}}
    \subfloat[after]{\includegraphics[scale=.450]{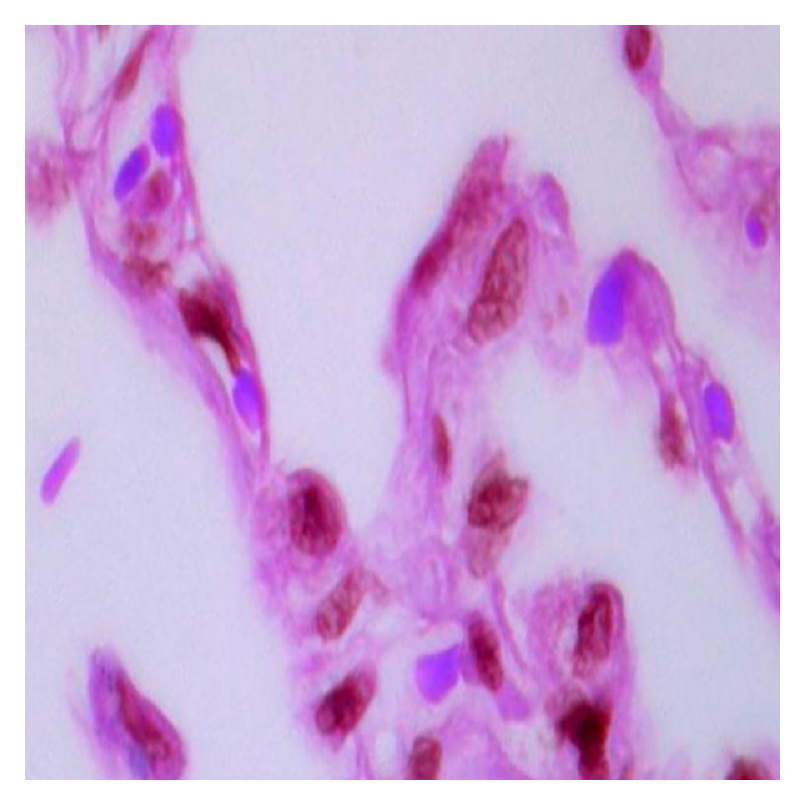}}
    
    \caption{Lung adenocarcinoma, squamous cell carcinoma, benign tissue before and after pre-processing}
	\label{fig:lung_pre-process}
	
\end{figure*}

After pre-processing, the resultant images are sharper, bright-colored, and more discernible details are visible than the original which are suitable to drive into the model and get better performance than others. Fig \ref{fig:lung_pre-process} depicts the before and after image pre-processing for lung cancer adenocarcinoma, squamous cell carcinoma, and benign tissue respectively, where the upper part (a,b,c) are before pre-processing and the lower part (d,e,f) are the after pre-processing of lung cancer image.  

\begin{figure*}[!htbp]
	\centering
	\subfloat[before]{\includegraphics[scale=.450]{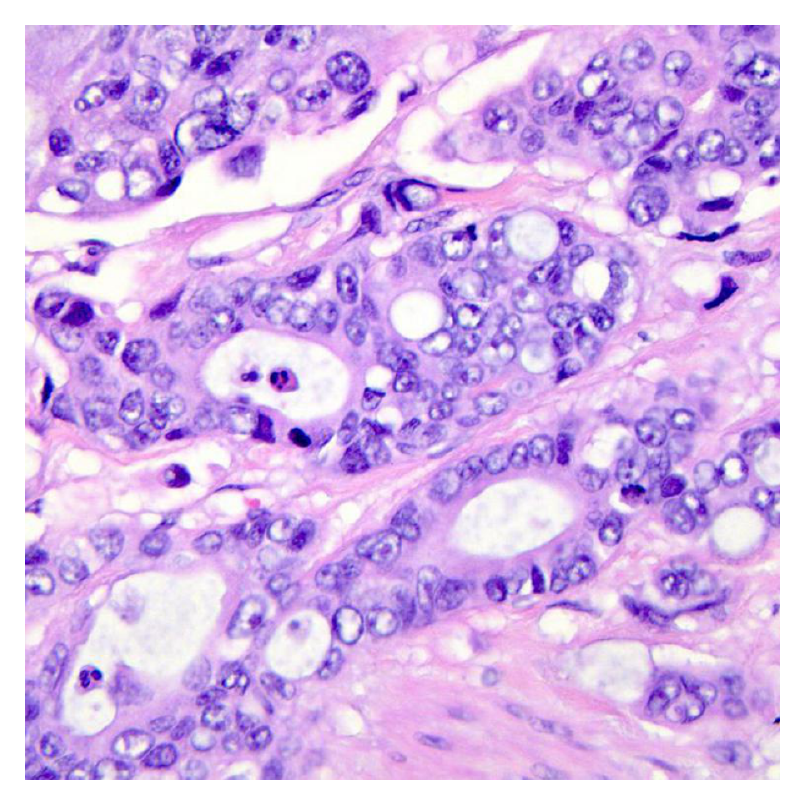}}
	\subfloat[after]{\includegraphics[scale=.450]{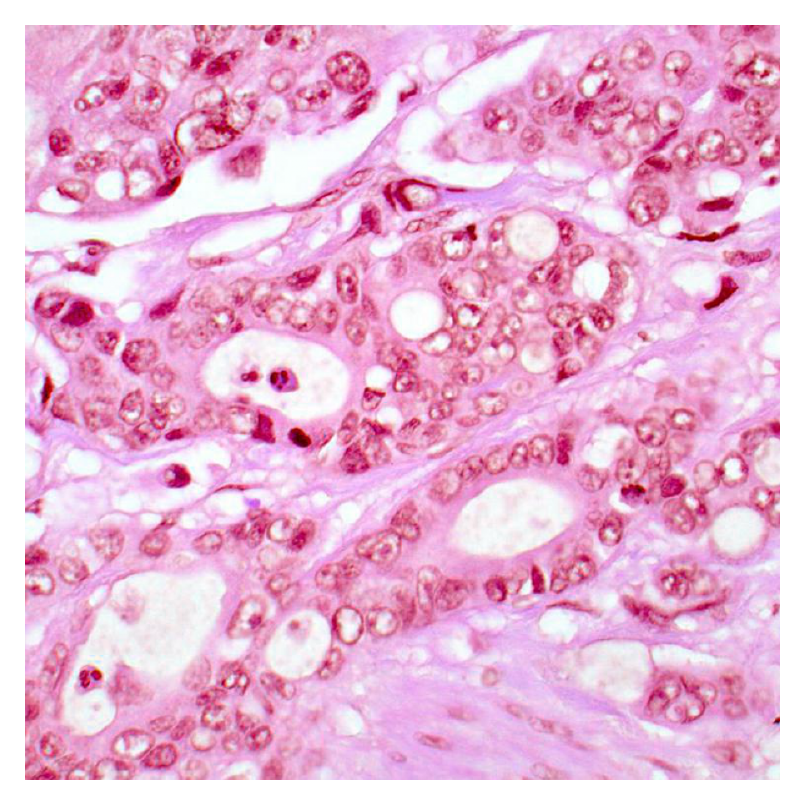}}
    \hspace{0.1cm}
	\subfloat[before]{\includegraphics[scale=.450]{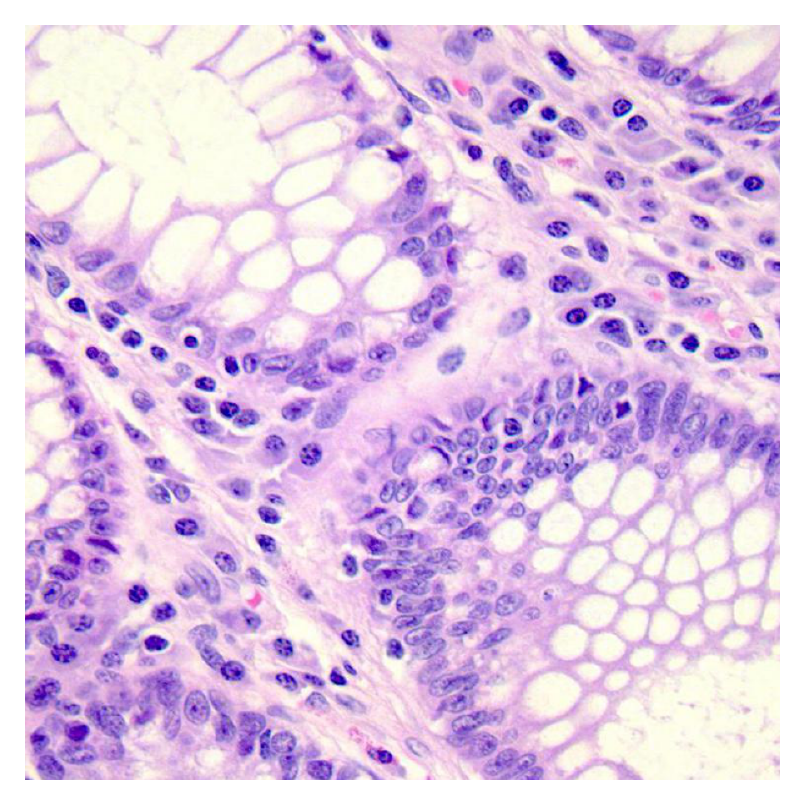}}
	\subfloat[after]{\includegraphics[scale=.450]{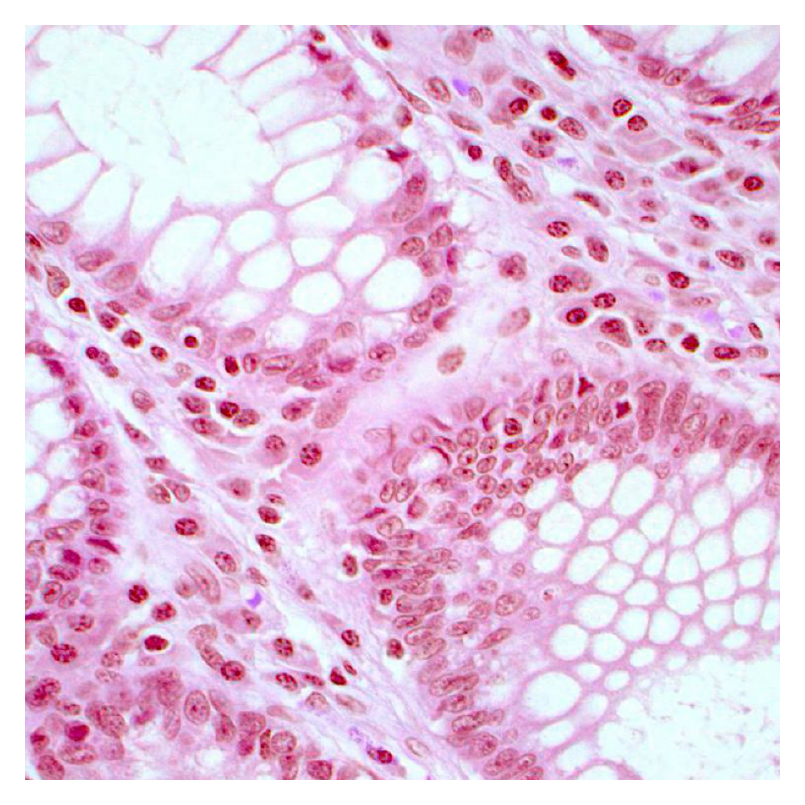}}
	\hspace{0.1cm}
	\caption{Colon adenocarcinoma and benign image before and after pre-processing}
	\label{fig:colon_pre-process}
\end{figure*}

Fig \ref{fig:colon_pre-process} depicts the before and after image pre-processing for colon cancer adenocarcinoma and colon benign tissue respectively, where a and b are the before and after pre-processing images for colon adenocarcinoma images and c and d are the before and after pre-processing image for colon benign tissue image.

\subsection{Deep Feature Extraction}

Transfer learning is the method of applying features learned on one problem to solve a new, related problem \citep{tf_learning}. This deep learning approach has received a lot of attention, so we adopt several existing models in our methodology and adjust it to fit our needs. Rather than training all of the model's layers, we isolate a few and use the trained weights in those frozen layers to derive certain properties from our datasets \citep{tf_from_scratch}. The significant properties of an image have a big impact on how well an identification operation works. In our paper, we analyze our datasets using five different TL models such as VGG16, VGG19, MobileNet, DenseNet169, and DenseNet201 to extract features in addition to assessing each individual's effectiveness by using various ML algorithms. 

\begin{itemize}
\item 
VGG16, a CNN model produced by the Visual Geometry Group (VGG) of the University of Oxford, was named the winner of the 2014 ILSVRC object identification algorithm \citep{karen_simonyan}. VGG16's main contribution is to show that increasing network depth can enhance network performance in some cases. VGG16 improves on the standard AlexNet by using numerous 33 convolution cores to substitute the larger convolution cores (1111, 707, 55), which can extend the depth of the network and minimize the number of network parameters \citep{yang2021novel}. There are 3 parts to VGG16 transfer learning such as convolution, pooling, and fully connected layers. Filters are employed in the convolution layer to extract information from images, and the kernel as well as stride  size are the most significant characteristics in this layer. To lessen the dimensionality and calculations, the poling layer is employed to shrink the network's spatial size. They have fully connected interactions with the previous layers in fully connected layers.
\item 
The VGG19 transfer learning model, which is also a CNN model, was introduced by \citep{simonyan2014very}. It is a 19 layers weighted ground feature extractor that is extensively utilized \citep{lin2021cotton}.  The consisting of it's layers is the combination of 16 convnet and 3 dense layers which are the linked layers to categorize pictures into 1000 item categories. The ImageNet repository was employed to develop VGG19, which comprises over a million pictures separated into 1000 classes. The sequence of the model is 2 conv 1 max.pool, 2 conv 1 max.pool, 4 conv 1 max.pool, 4 conv 1 max.pool, 4 conv 1 max.pool and 3 FC layers. Due to the employment of numerous 3x3 filters within every convnet, it is a widely common picture predictive model.
\item  
One of the Convolutional networks that have already been trained is MobileNet \citep{howard2017mobilenets}. The quantity of parameters that have been lower than others distinguishes MobileNet from other existing models. It is built on depthwise separable convolutions, which were developed in \citep{sifre1687rigid} to reduce the computing effort in the initial few layers, depthwise and pointwise convolutions are two layers by which it is constituted. The technique of filtering the input without creating additional features is known as depthwise convolution. As a result, the pointwise convolution procedure, which generates innovative features, was integrated. Thus, depthwise separable convolution was coined to describe the merging of two layers. It assigns a unique filter for each stream of input using depthwise convolutions and afterwards utilizes 1x1 conv (pointwise) for constructing a joint distribution of output from the depthwise layer. Batch Normalization (BN) with Rectified Linear Unit (ReLU) are employed after each convolution. It's efficiency was determined by a trade-off between delay and correctness. Due to the compactness, we can perform our tasks using this method on the CPU in lieu of GPU. It can be used for categorization, identification, and segmentation in the same way as previous large-scale models that have been released \citep{sae2019convolutional}. 
\item
DenseNet is a Dense Convolution Neural Network, introduced by \citep{huang2017densely}. In DenseNet, each layer influences all other layers in a feed-forward manner [30]. There are five levels in a DenseNet block. The very first four layers are dense, and the final layer is a transition layer. This dense block has a growth rate of (k) for each layer. If the growth rate is 4 then  a 2x2 average pool with a 1x1 conv and a stride of 2 exists in the transition layer. There are 1x1 and 3x3 conv with a stride of 1 in the dense layer. Two popular DenseNet networks such as DenseNet169 and DenseNet201 are utilized in our research experiment.
\end{itemize}

\begin{figure*}[!h]
\centering
\subfloat[The original VGG16 model]{\includegraphics[scale=.26]{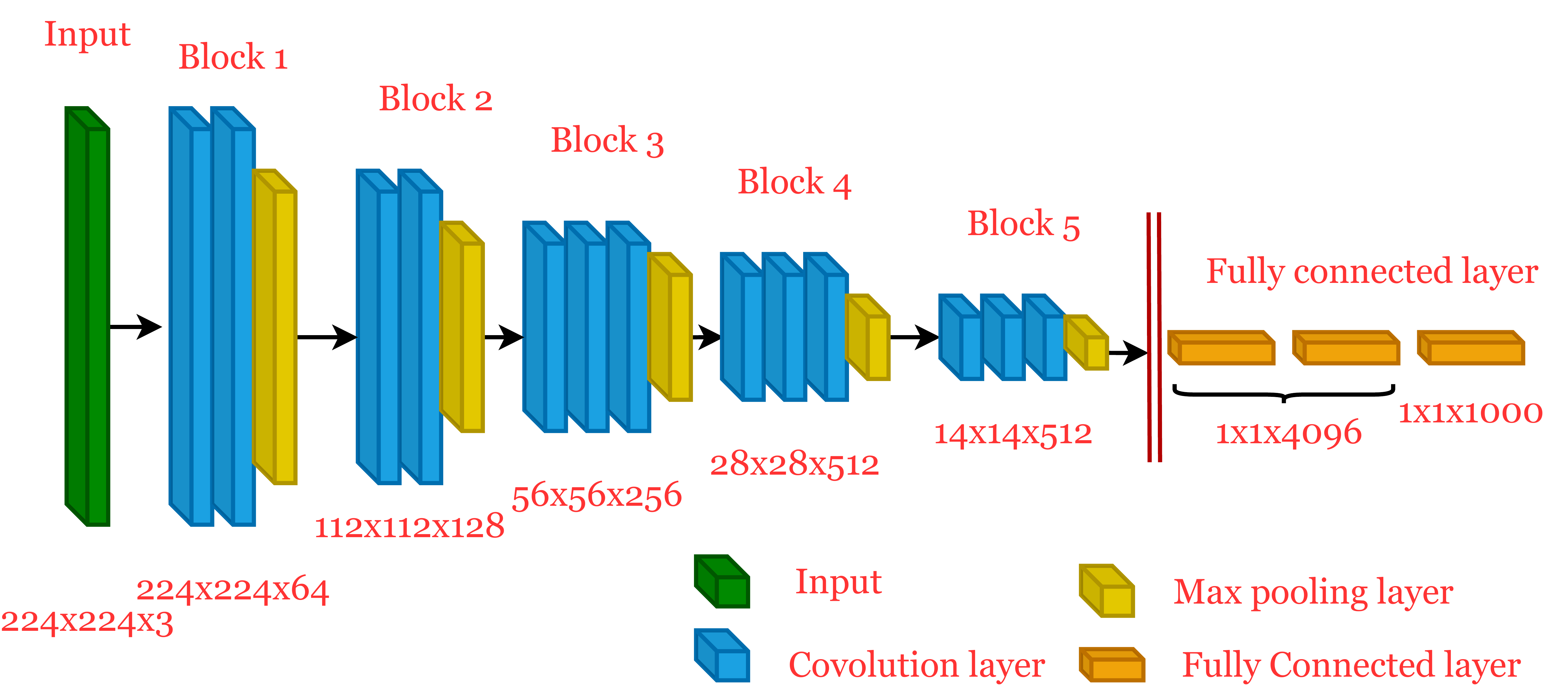}} \hspace{0.1cm}
\subfloat[Deep feature extracted VGG16 model]{\includegraphics[scale=.18]{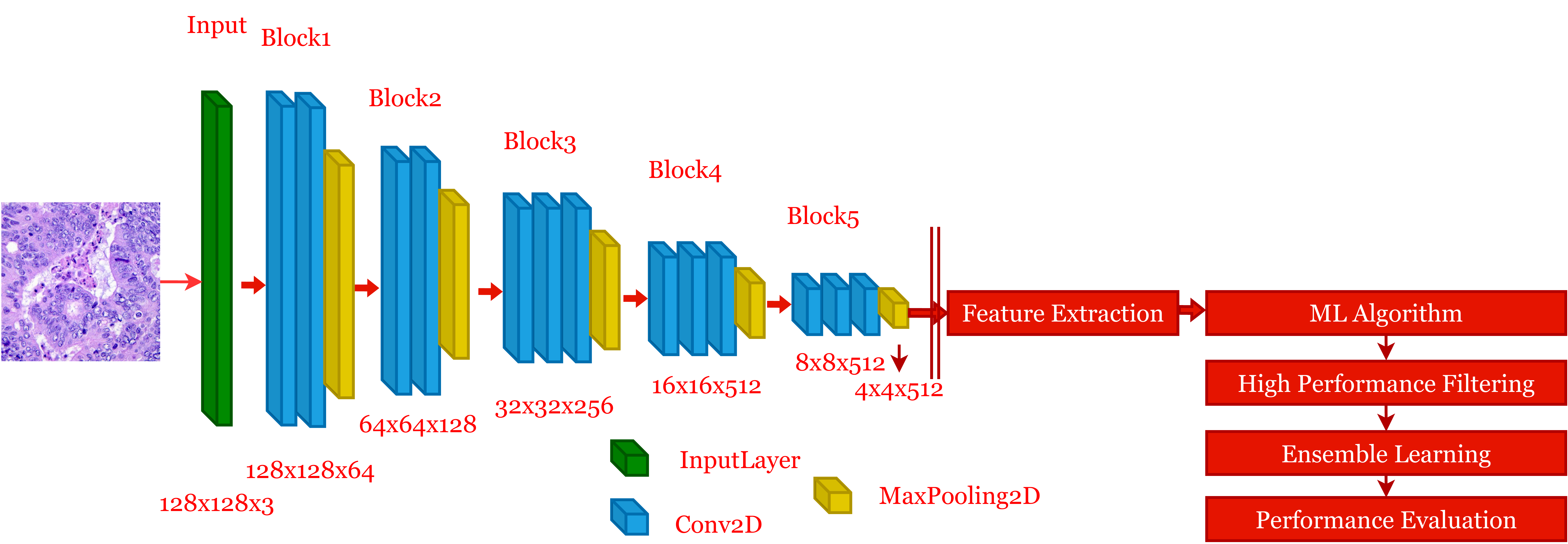}}\hspace{0.1cm}
\caption{A sample architecture of original VGG16 model and deep feature extracted VGG16 model }
\label{fig:feature_extract}
\end{figure*}

Fig. \ref{fig:feature_extract} illustrates a sample of the feature extraction procedure where it includes the original architecture of VGG16 and deep feature extraction from the VGG16 model. The composition of the original VGG16 CNN model has 13 convolutional (conv) layers, 3 fully connected (FC) layers, and 5 pooling layers. The sequence of the model is 2 conv 1 pool, 2 conv 1 pool, 3 conv 1 pool, 3 conv 1 pool, 3 conv 1 pool and 3 FC layers. It employs a small kernel (3x3) with 1 stride on all conv layers. The max pool layer continually derives after the conv layers. It takes 3 channels of images with a resolution of 224x224. In the 3 FC layers, the first two FC layers have an identical channel size of 4096 and the last one has a channel size of 1000, which defines the number of class labels in the ImageNet data. The soft-max layer is the output layer, and it's in charge of the input image's assigned probability. On the other hand, the deep feature extraction from the VGG16 model is the same as the original VGG16 model except for the last 3 FC layers. Without involving the last 3 FC layers, we have just extracted features from this model and grazed them into different ML models after that we apply HPF and get the top 3 filtered ML algorithms and then apply ensemble learning with these algorithms and evaluate the performance. The same procedure is applied to all of the TL models to extract features.

\begin{table}[]
\centering
\begin{tabular}{lll}
\hline
 Dataset &  Before feature extraction &  After feature extraction \\ \hline
 Lung &  4200, 128, 128, 3 &  4200, 16384 \\ 
 Colon &  2800, 128, 128, 3 &  2800, 16384 \\ 
 Cancer(lung and colon) &  10000, 128, 128, 3 &  10000, 16384 \\ \hline
 
\end{tabular}%
\caption{Shape of feature}
\label{tab:fs}
\end{table}

In this feature extraction process, we fixed the last 3 FC layers and extracted the features from the last block of the max-pooling layer. In the feature extraction process, the before and after feature shape is shown in \ref{tab:fs} where the input size of the image is 128*128*3. The final feature size after feature extraction is 16384. Hence, before feature extraction the size of features is n*128*128*3, and after extraction n*16384. Here, n is the number of input images. So, for lung cancer, the input feature size is 4200*128*128*3 and the output feature size is 4200*16384, for colon cancer, the input feature size is 2800*128*128*3, and the output feature size is 2800*16384, for (lung and colon) cancer, the input feature size is 10000*128*128*3 and output feature size is 10000*16384. After that, we provision the extracted features into the ML algorithms to assess the efficiency of each machine learning model. This procedure is applied for all of the above described TL models.
 
\subsection{Machine Learning Algorithms}

In this section, we have described several ML algorithms which is used in our model to evaluate the performance. 

\begin{itemize}
    \item Random Forest (RF) is a meta-approximation that employs averaging to enhance accuracy level and multiple decision tree classifiers are fitted to various sub-trials of the dataset to prevent over-fitting \citep{alkhatib2020predictive}. It is a combination of several uncut DTs derived from different bootstrap samples of the training data and each attribute subset is sampled separately from the actual feature space \citep{breiman2001random}. It creates several independent decision trees from the training dataset's original features, then votes to combine all of the trees into a single classification model \citep{pan2014predicting}. It achieves a better prediction accuracy due to employing the outcomes of numerous decision trees to construct a prediction \citep{ahsan2021enhancing}.

    \item Support Vector Machine (SVM) is a classification procedure initiated with the detection of the hyperplane and then the classification of the classes \citep{safaldin2021improved}. For classification analysis, SVM is a popular machine learning tool. SVM increases generalization by reducing the learning machine's structural risk. SVM has been used in many contexts because of its high recognition rate and fast training speed when classifying relatively small datasets \citep{shen2016evolving}. It has been shown to be an efficient strategy for classification tasks because it adheres to the proper risk management design to maximize generalizability. Furthermore, because solving an SVM is effectively a convex optimization problem, the procured internally optimal solution must also be a globally optimal one. As a result, SVM is used to train the base classifiers in this case \citep{gu2019novel}. 

    \item Logistic Regression (LR) is a conventional parametric approach that is a generalized linear model extension that has been typically employed in diverse areas, including health care \citep{yang2014statistical}. It uses a logit relation equation to predict the output. Using backward/forward elimination with a p-value less than 0.05, we can pick the most significant predictor parameters using the method's iterative selection approach. The Realimpo package in R 3.3.1 statistical program is used to measure the relative value of each variable \citep{gromping2006relative}. It conjectures to find out how to generate a categorical output from input variables and also forecasts a categorical dependent variable's output. So, the final output must be either categories or discrete.
    
    \item Multilayer Perceptron (MLP) is a common ANN architecture that consists of a series of layers made up of neurons as well as their connections. It can measure the weighted sum of its inputs before applying an activation function to generate a signal that will be sent to the next neuron \citep{castro2017multilayer}. Between the input and output layers, it has one or more hidden layers. The neurons are arranged in layers and connections are often guided from lower to upper layers, and the neurons in the same layer are not linked\citep{ramchoun2016multilayer}. In the input layer, the number of neurons equals the number of measurements for the pattern query, while the number of neurons in the output layer equals the number of classes.

    \item Extreme Gradient Boosting (XGB) is a supervised ML technique that improves speed and performance by using gradient boosted decision trees. It is fast as compared to other implementations of gradient boosting \citep{chen2015higgs}. The final prediction is made by a combination of the residuals of prior models which are predicted from the generated novel models. It uses a gradient descent strategy to reduce loss and to boost up the model for improved performance. It's a scalable end-to-end tree boosting method that data scientists utilize to obtain cutting-edge outcomes on a variety of machine learning problems \citep{chen2016xgboost}. The objective functions in XGboost consist of two parts training loss and regularization with $\theta$ the optimal settings for $x_i$ training data and $y_i$ labels. Equ. \ref{equ:loss} Shows objective functions of XGBboost.
    
    \begin{equation}
        O(\theta ) = L(\theta ) + \Omega (\theta )
        \label{equ:loss}
    \end{equation}
     Here, $L$ stands for the training loss function, which indicates how well our model predicts the training datasets. 
         
    \item Light Gradient Boosting (LGB) is a new gradient learning paradigm based on the decision tree idea \citep{ke2017lightgbm}. It outperforms XGBoost, is distinguished by its use of less storage, its use of a leaf-wise growth technique with depth limits, and its use of a histogram-based algorithm to speed up the learning process \citep{yang2020comparative}. LightGBM builds a k-width histogram by discretizing continuum floating-point eigenvalues into k bins employing the aforementioned histogram algorithm. Moreover, the histogram approach does not necessitate extra archiving of pre-sorted findings and results can be saved in an 8-bit integer upon feature determination, which reduces storage use to 1/8 of what it was before. In spite of this, the model's efficiency suffers because of the strict dividing mechanism \citep{mohammadi2021modeling}.
\end{itemize}

\subsection{High-Performance Filtering}
 
High-Performance Filtering (HPF) is a process in which we have selected the top three ML algorithms to feed into our ensemble learning process. The output of this HPF process is to find the estimators which we can feed into our ensemble learning model to generate our final output. To optimize the consistency and accuracy of estimates, we use the best three algorithms in the ensemble learning process. In some circumstances, if we acquire worse accuracy in half of the six ML algorithms, the ensemble learning models would not outperform the base learners. Moreover, if we select the top two ML algorithms the robustness is comparatively legislation. Hence, to circumvent these issues, we select the top three or half of the trained ML algorithms. Moreover, the model's training time will be comparatively less than the top 4 as it used only top 3 and the accuracy rate will be comparatively higher than the top 2 as it used top 3. 
 
In the following section, we have tabulated the feature extraction performance of our experiments  as well as visualized the accuracy performance in the graphical form.
In Table \ref{tab:lung_acc_analysis}, \ref{tab:colon_acc_analysis} and \ref{tab:cancer_acc_analysis} show that among all the algorithms the SVM, LR, and MLP provide the better performance.

\begin{itemize}
\item 
    For \ref{tab:lung_acc_analysis} table, the average accuracy rate is 94.95\%, 97.67\%, 97.67\%, 97.67\%, 95.86\%, 96.76\% for RF, SVM, LR, MLP, XGB, and LGB. Among all SVM, LR, and MLP give the highest average accuracy than others. Hence, we select these top three algorithms for lung cancer to fit into the ensemble models and further analysis.

    \begin{table*}[]
    \centering
    \begin{tabular}{lllllll}
    \hline
    Algorithm & VGG16 & VGG19 & MobileNet & DenseNet169 & DenseNet201 & Average \\ \hline
    RF & 93.57 & 94.05 & 95.71 & 94.52 & 96.9 & 94.95 \\ 
    SVM & 96.9 & 97.62 & 98.57 & 97.14 & 98.1 & 97.67 \\ 
    LR & 96.9 & 96.67 & 98.81 & 97.14 & 98.33 & 97.57 \\ 
    MLP & 96.9 & 96.67 & 98.1 & 97.62 & 99.05 & 97.67 \\ 
    XGB & 94.05 & 95.71 & 96.19 & 95.95 & 97.38 & 95.86 \\ 
    LGB & 95.24 & 96.43 & 97.38 & 96.67 & 98.1 & 96.76 \\ \hline
    \end{tabular}%
    \caption{Accuracy performance of ml algorithms on TL for lung cancer (in \%)}
    \label{tab:lung_acc_analysis}
    \end{table*}
\item
    For \ref{tab:colon_acc_analysis} table, the average accuracy rate is 98.29\%, 99.57\%, 99.57\%, 99.57\%, 97.79\%, 98.64\% for RF, SVM, LR, MLP, XGB, and LGB. Among all SVM, LR, and MLP give the highest average accuracy than others.  Hence, we select these top three algorithms for colon cancer to fit into the ensemble models and further analysis.
    
    \begin{table*}[]
    \centering
    \begin{tabular}{lllllll}
    \hline
    Algorithm & VGG16 & VGG19 & MobileNet & DenseNet169 & DenseNet201 & Average \\ \hline
    RF & 97.5 & 97.14 & 99.29 & 98.57 & 98.93 & 98.29 \\ 
    SVM & 99.64 & 99.29 & 100.0 & 99.29 & 99.64 & 99.57 \\ 
    LR & 99.64 & 99.29 & 100.0 & 99.29 & 99.64 & 99.57 \\ 
    MLP & 99.64 & 99.29 & 100.0 & 99.29 & 99.64 & 99.57 \\ 
    XGB & 95.36 & 97.5 & 98.93 & 98.21 & 98.93 & 97.79 \\ 
    LGB & 97.14 & 97.86 & 99.64 & 99.29 & 99.29 & 98.64 \\ \hline
    \end{tabular}%
    \caption{Accuracy performance of ml algorithms on  for colon cancer (in \%)}
    \label{tab:colon_acc_analysis}
    \end{table*}

\item
    For \ref{tab:cancer_acc_analysis} table, the average accuracy rate is 95.8\%, 98.54\%, 98.42\%, 98.58\%, 96.70\%, 97.28\% for RF, SVM, LR, MLP, XGB, and LGB. Among all SVM, LR, and MLP give the highest average accuracy than others.  Hence, we select these top three algorithms for (lung and colon) cancer to fit into the ensemble models and for further analysis.
\end{itemize}

\begin{table*}[]
\centering
\begin{tabular}{lllllll}
\hline
Algorithm & VGG16 & VGG19 & MobileNet & DenseNet169 & DenseNet201 & Average \\ \hline
RF & 95.2 & 94.7 & 96.8 & 95.5 & 96.8 & 95.8 \\ 
SVM & 97.9 & 97.8 & 99.4 & 98.3 & 99.3 & 98.54 \\ 
LR & 98.0 & 97.6 & 99.2 & 98.1 & 99.2 & 98.42 \\ 
MLP & 98.4 & 97.9 & 99.0 & 98.3 & 99.3 & 98.58 \\ 
XGB & 95.6 & 95.6 & 97.0 & 97.5 & 97.8 & 96.7 \\ 
LGB & 96.1 & 96.2 & 98.5 & 97.7 & 97.9 & 97.28 \\ \hline
\end{tabular}%
\caption{Accuracy performance of ml algorithms on TL for lung and colon cancer (in \%)}
\label{tab:cancer_acc_analysis}
\end{table*}

\begin{figure*}[!htbp]
	\centering
	\subfloat[lung cancer]{\includegraphics[scale=.400]{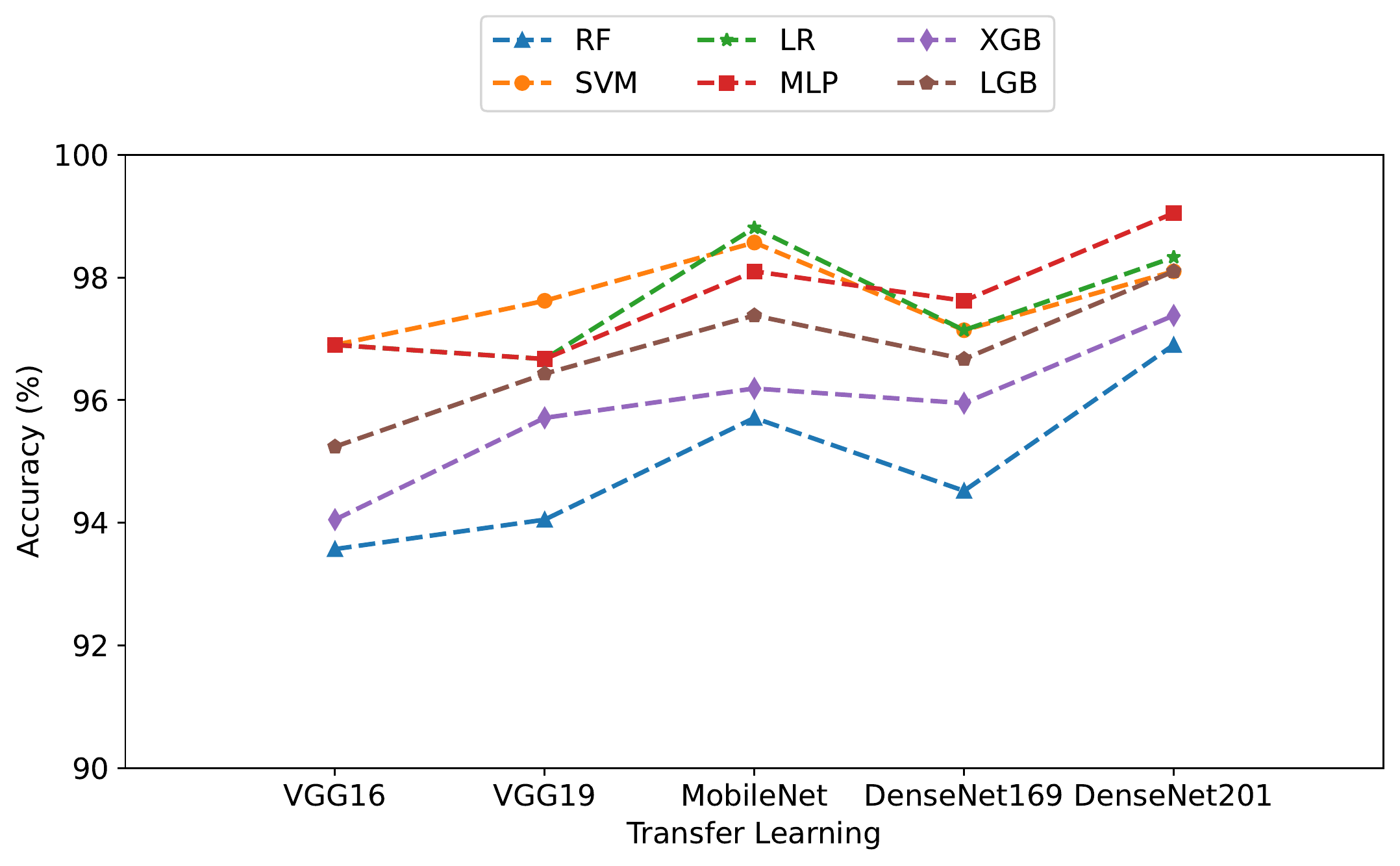}}
	\subfloat[colon cancer]{\includegraphics[scale=.400]{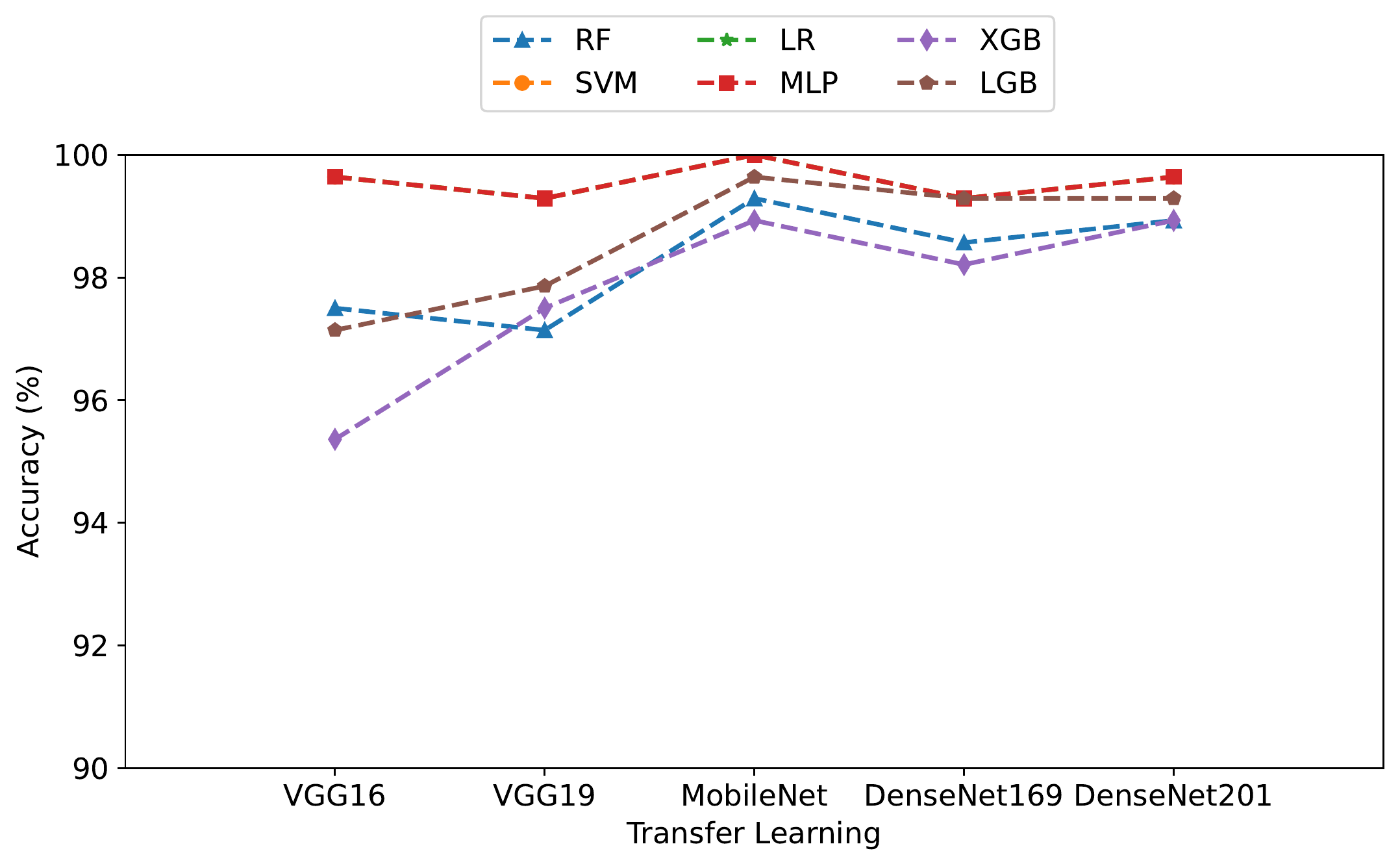}} \hspace{0.1cm}
	\subfloat[lung and colon cancer]{\includegraphics[scale=.450]{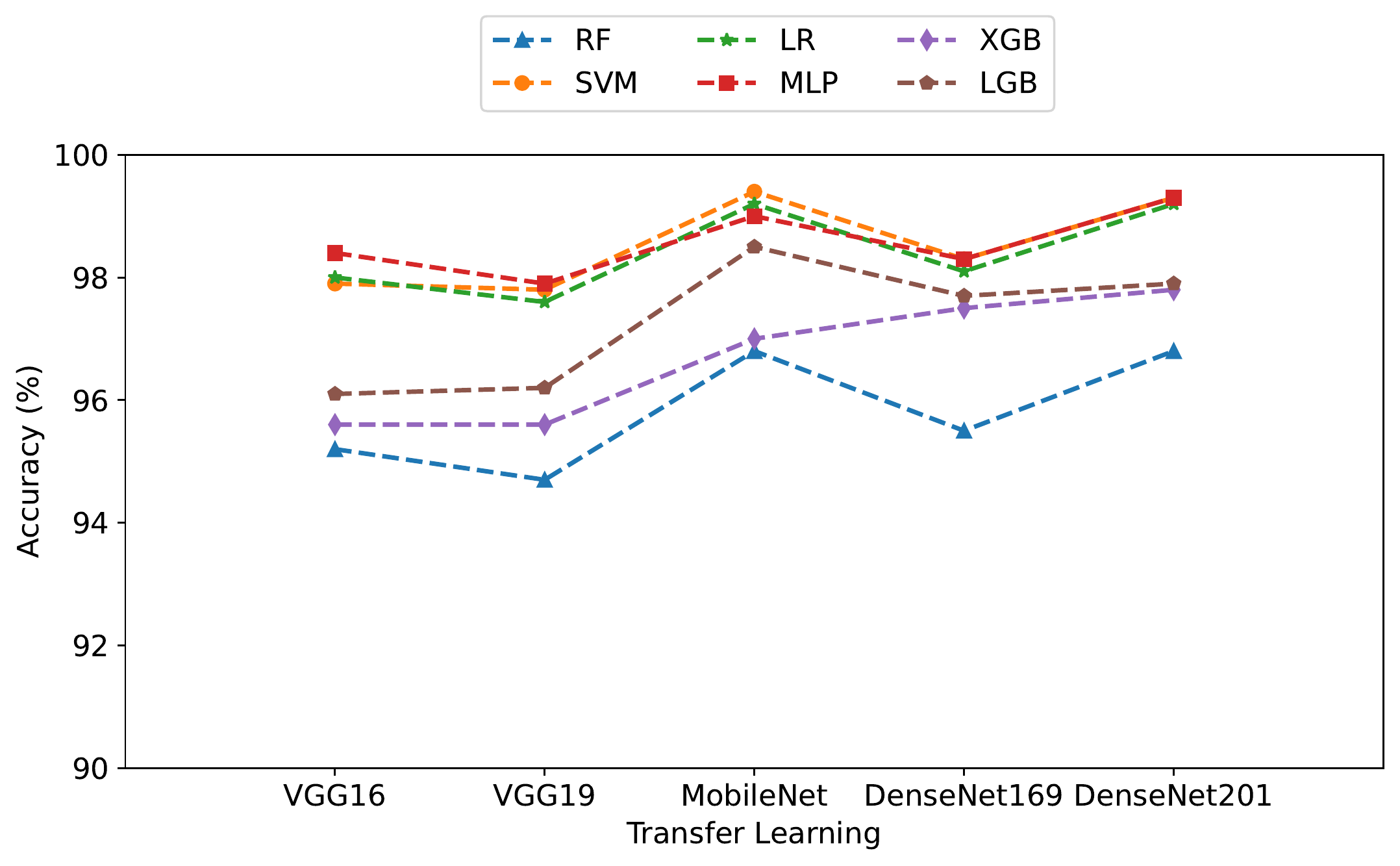}} 
	\caption{Accuracy performance analysis of ML algorithms on TL for lung and colon cancer}
	\label{fig:acc_analysis}
\end{figure*}

Fig \ref{fig:acc_analysis} illustrates that for lung, colon, and lung and colon graphs, among all the ML algorithms SVM, LR, and MLP outperform for each TL model. As a result, we select the most performance gained algorithms from all ML algorithms such as SVM, LR, and MLP for lung, colon and 
lung and colon cancer. 

After successfully filtering the high-performance ML algorithms, we fit these into our ensemble model to get better performance. For each dataset, we fit those selected top 3 selected ML algorithms and fit into the two ensemble voting classifiers namely majority (hard) and weighted average (soft). After that, we assess the effectiveness of our models for each dataset. 

\subsection{Ensemble Learning}
Ensemble learning (EL), also known as multiple classifier or committee-based learning, is a way of training and integrating numerous learners to successfully deal with a problem \citep{zhou2016machine}. During the last few generations, it has sparked a lot of attention, particularly in the AI and machine learning areas. It is not only justified and fairly efficient, but it is also incredibly adaptable in a broad variety of essential areas and contributes significantly, so this attention is pretty well deserved \citep{dietterich2002ensemble}. Regarding majority rule voting, hard employs anticipated class labeling. Soft estimates each class label using the argmax where the intended function's maximum value is returned, of the summing of the estimated probabilities, which is ideal for just a group of well-calibrated learners \citep{featurest}.

In soft voting, we forecast the class labels depending on the classifier’s estimated probabilities \citep{kumari2021ensemble}, where we used averaging the class probability \citep{raschka2015python}. In the implementation by default, it calculates the soft voting using a uniform value. Hence, the weight that is used in the weighted average process is a uniform value. Moreover, it has been used since the classifiers that we have used are well-calibrated which means the scores of the classifiers are in good agreement with probability estimations. The way of the soft voting procedure is shown in the following equation.

\[\hat y = \arg max \sum\limits_{j = 1}^n {({w_j}{p_j})} \]
where, $w_{j}$ is the weights of classifiers and $p_{j}$ is the prediction of classifiers.

\begin{figure*}[!htbp]
\centering
  \includegraphics[width=0.55\textwidth]{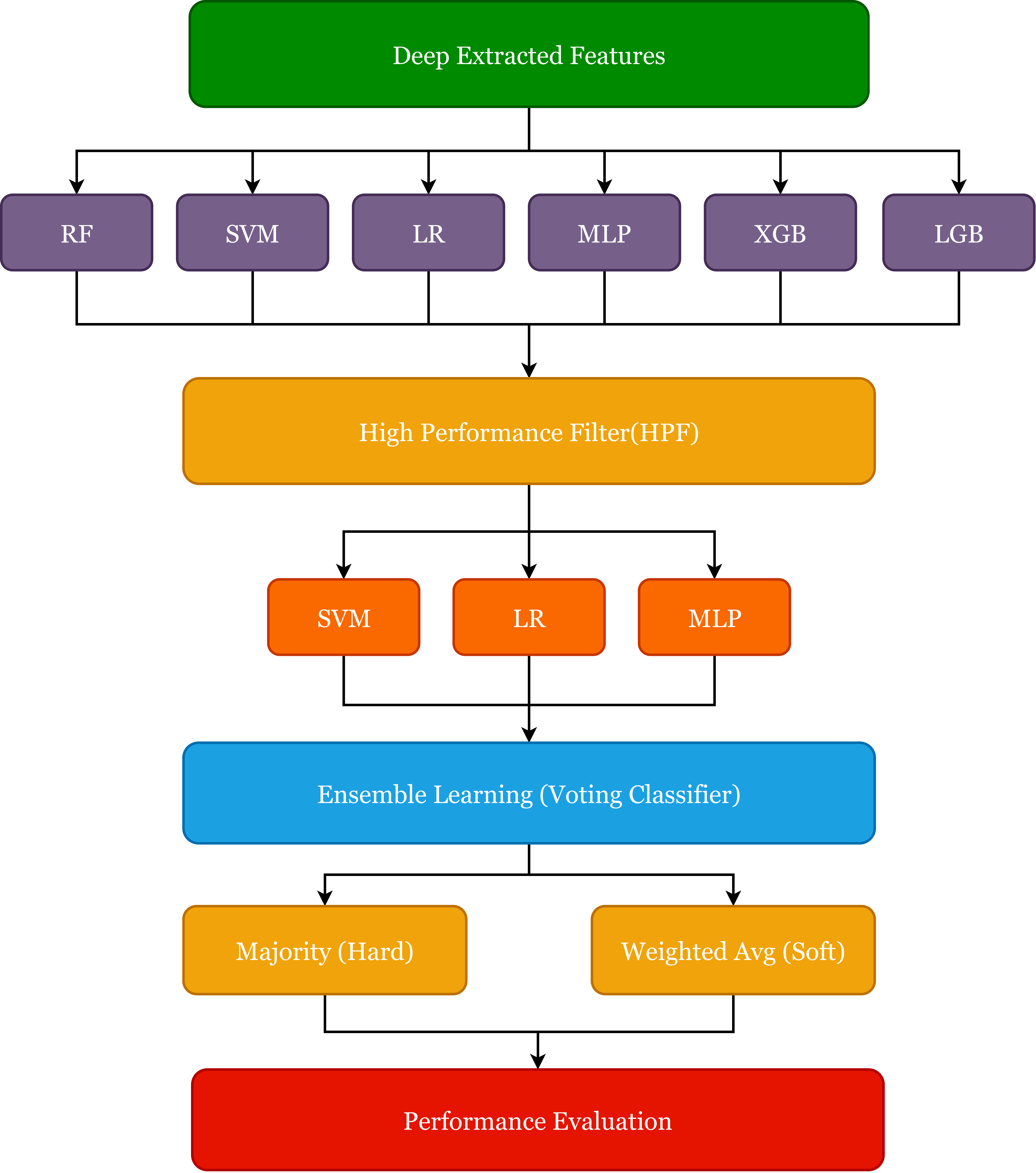} 
\caption{Ensemble learning approach with HPF}
\label{fig:ensemble}
\end{figure*}

Fig. \ref{fig:ensemble} illustrates the EL process with HPF, where the deep extracted features are fed into the different ML algorithms, and then by applying HPF we get three algorithms; after that, we apply EL to these filtering algorithms. In EL, the soft and hard voting algorithms are utilized to evaluate the performance. 
After getting the estimators from HPF, we use ensemble learning with the HPF algorithm. Where we used our estimators with hard and soft for hard and soft voting classifiers respectively.

\section{Experimental setup and performance metrics}
\label{sec:experimental}
The experiments are carried out on a machine running Microsoft Windows 10 Pro, with an Intel(R) Core (TM) i3-6006U CPU running at 2.00GHz, 2000Mhz, 2 cores, 4 logical processors, and 120GB SSD, 1TB HDD, 8GB RAM and 26GB Virtual Memory. The experiment is carried out in Anaconda Navigator using a Jupyter notebook. To implement the suggested model, the Python programming language and various common libraries such as pandas, NumPy, Matplotlib, Seaborn, TensorFlow, Keras, Scikit-learn, and others are utilized.
\\
Numerous performance metrics appraise our proposed model, including accuracy, recall, precision, f1-score, ROC curve, confusion matrix, MSE, MAE, and RMSE. The metrics for appraising the performance are formulated in the following:

\begin{itemize}
    \item The Confusion Matrix is a technique to assess machine learning classification efficiency. It is a table-like structure with four (TP, TN, FP, FN) combinations of predicted and actual values. Table \ref{sec:conclusion} shows a confusion matrix where the TP (True Positive) denotes a correctly anticipated positive value, TN (True Negative) denotes a correctly projected negative value, FN (False Positive) gives an inaccurately forecasted positive value, and FN (False Negative) indicates an incorrectly predicted negative value. It is very useful in helping to determine accuracy, recall, precision, f1-score, and ROC curve. 
    
    \begin{table}[!h]
        \centering
        \begin{tabular}{lll}
        \hline
            & Actual   positive & Actual   negative \\ \hline
            Predicted   positive & TP & FP \\ 
            Predicted   negative & FN & TN \\ \hline
        \end{tabular}
        \caption{Confusion Matrix}
        \label{table:confusion}
    \end{table}
    
    \item Accuracy is the most obvious performance metric, and it is directly proportional of correctly predicted observations to total observations.
    \begin{equation}
    Accuracy =\frac{TP+TN}{TP+FP+FN+TN}
    \end{equation}
    
    \item The ratio between the correctly predicted positive value and the total number of predicted positive values is defined as precision. It is depicted as:
    \begin{equation}
    Precision=\frac{TP}{TP+FP}
    \end{equation}
    \item The ratio between the correctly predicted positive value and all numbers of actual values is defined as recall. It is depicted as:
    \begin{equation}
    Recall=\frac{TP}{TP+FN}
    \end{equation}
    
    \item The harmonic means of precision and recall values for a classification problem is defined as f1-score. f1-score is depicted as:
    \begin{equation}
    F1-score=2*\frac{(precision *recall)}{(precision +recall)}    
    \end{equation}
    \item The mean absolute error (MAE) is a metric for comparing errors within related measurements that describe a similar occurrence. It is an absolute inaccuracy of the arithmetic average of predicted and actual values \citep{willmott2005advantages}.
    \begin{equation}
            MAE = {\frac{{\sum\limits_{i = 1}^n pred(i) - act(i)}}{n}}
    \end{equation}   
    
    \item The mean squared error (MSE) is a statistic that calculates the mean of the squared residuals, or the mean squared discrepancy between actual and forecast values. Because of variability or if the predictor does not allow for data that could generate a good prediction, the most important fact about MSE is that it is usually typically absolutely positive (rather than zero) \citep{lehmann2006theory} \citep{das2004mean}. 
    \begin{equation}
            MSE = {\frac{{\sum\limits_{i = 1}^n (pred(i) - act(i))^{2}}}{n}}
    \end{equation}
    
    \item The most commonly used evaluation metric in regression problems is the root mean square error (RMSE). It is premised on the notion that error is unbiased and seems to have a normal distribution. Outlier attributes have quite a massive effect on RMSE.  \begin{equation}
    RMSE = \sqrt{{\frac{{\sum\limits_{i = 1}^n   (pred(i) - act(i))^{2}}}{n}}}
    \end{equation}
    where pred(i) is the predicted value of i, act(i) is the actual value of i, and n is the total number of values.
    
    \item ROC curves are two dimensional plots that are often used to analyze and examine the effectiveness of classifiers \citep{fawcett2004roc}. ROC graphs clearly demonstrate a classifier's sensitivity or specificity exchange for all potential classification thresholds, enabling the grading and picking of classifiers based on particular user requirements, which are commonly linked with variable error costs and accuracy expectations \citep{krzanowski2009roc} \citep{vergara2008star}. AUC reflects the level of distinction while ROC is indeed a likelihood curve. It indicates how well the model can discriminate distinct categories. On the y-axis is the true positive rate, while on the x-axis is the false positive rate. AUC approaching 1 indicates that a predicted model is good at separability amongst class labels whereas approaching 0 indicates a lousy predicted model. In actuality, lousy signifies that the consequence is being mirrored \citep{narkhede2018understanding}. It's a strategy for visualizing the classification's efficiency \citep{gorunescu2011data}. Models with higher ROC curves are superior classifiers \citep{yulianto2019improving}.
    
    \item K-fold Cross-Validation (CV) is a basic procedure for dividing a training set into k smaller sets. The strategy for each of the k "folds" is to utilize the folds as training data for a model, which is then verified using the remaining data, and the average of the values computed in the loop is then included as an evaluation metric using k-fold cross-validation.
    
    Our cancer detection trials were carried out using a k-fold CV with a value of k=10. The dataset is partitioned into 80\% for training and the rest of the 20\% for testing in k-fold, and the dataset gets folded into 10 separate folds, each of which is used as a testing component throughout the folding process. The procedure of k-fold cross-validation is depicted in Fig. \ref{fig:kfld}.
    
    \begin{figure*}[!htbp]
    \centering
      \includegraphics[width=0.75\textwidth]{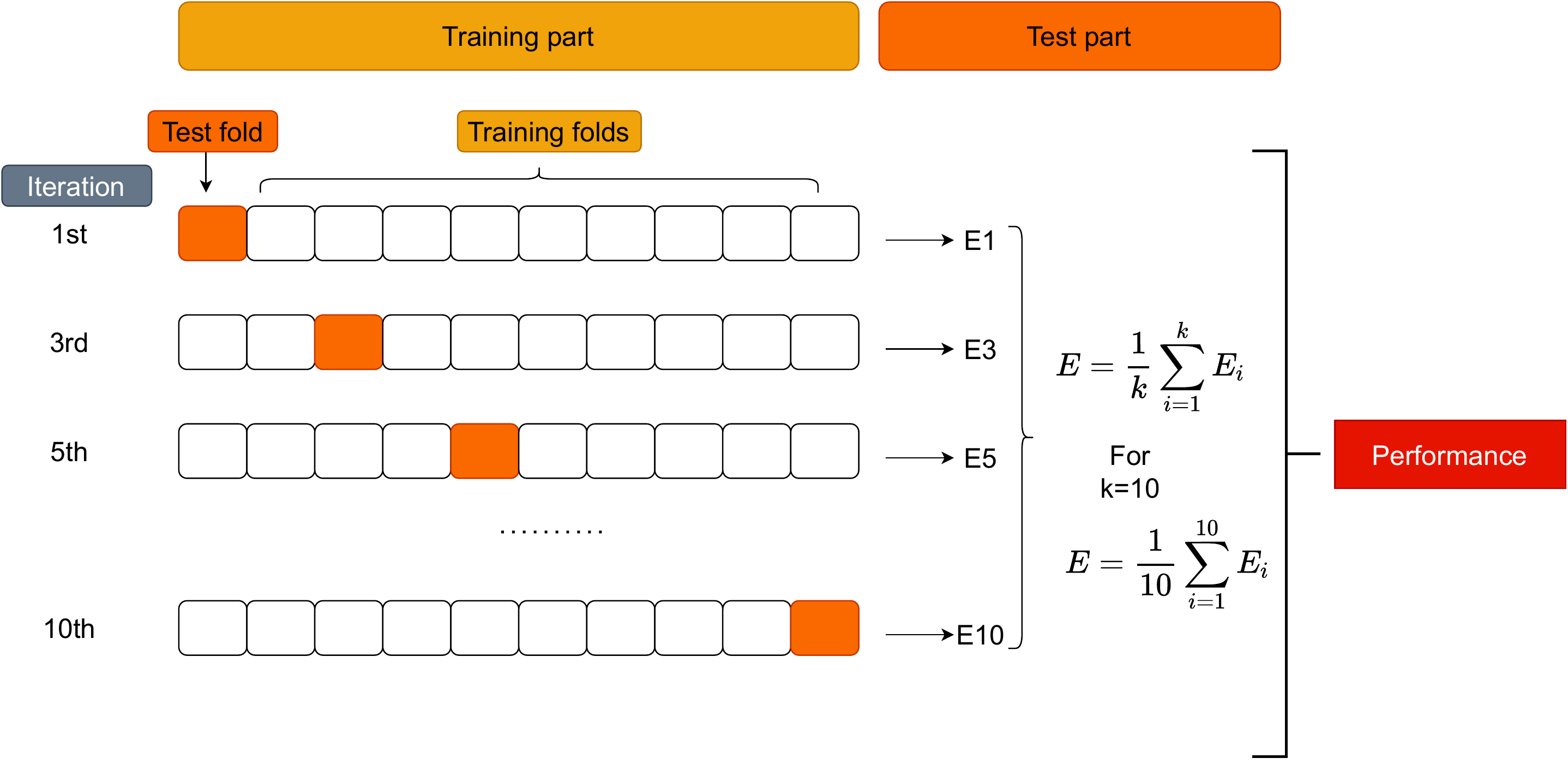}
    \caption{k-fold cross-validation process}
    \label{fig:kfld}
    \end{figure*}

\end{itemize}

\section{Performance Evaluation and Analysis}
\label{sec:evaluation}
This section analyzes the performance of different TL methods, followed by the analysis of our hybrid model based on the selected high-performance TL model and voting classifier model. We have found that among all TL models MobileNet model outperforms other models, and the soft voting classifier gives the highest performance than the hard voting classifier.  

\subsection{Performance evaluation of lung cancer}
\begin{figure*}[!htbp]
	\centering
	{\includegraphics[scale=.520]{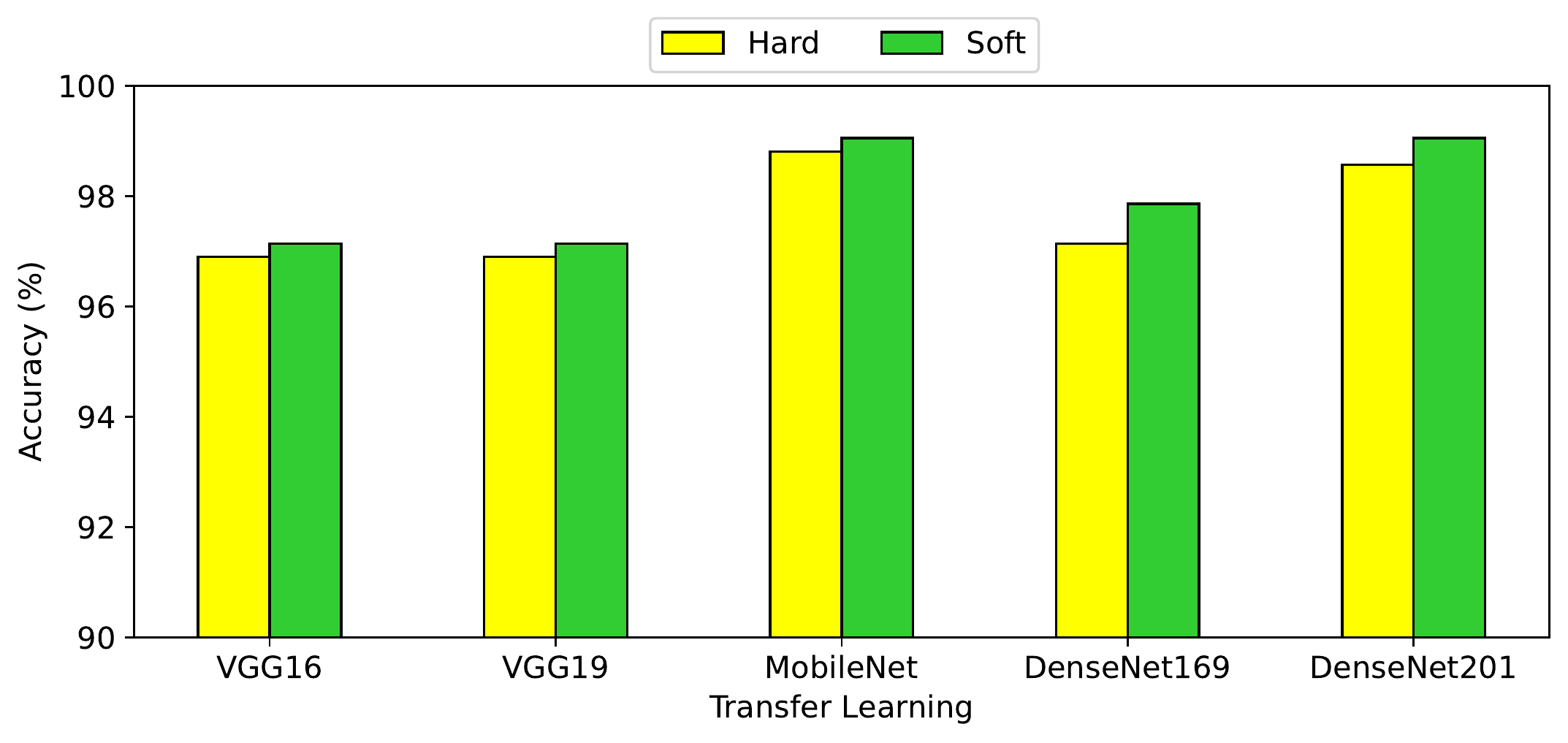}} \caption{Accuracy performance analysis for ensemble models on lung cancer dataset}
	\label{fig:lung_en_graph}
\end{figure*}

Table \ref{tab:lung_en_table} and Fig. \ref{fig:lung_en_graph} illustrate the performance of different TL models based on hard and soft voting in tabular format and visual graphical bar chart respectively. From Table \ref{tab:lung_en_table} we can see that the average accuracy rate for hard and soft voting classifiers is 97.66\% and 98.05\% respectively. Moreover in Fig. \ref{fig:lung_en_graph} for the graphical bar chart, we can see that the graphical upper bar is soft and lower is hard voting classifier, where for VGG16 96.9\%, 97.14\%; for VGG19 96.9\%, 97.14\%; for MobileNet 98.81, 99.05\%; for DenseNet169 98.57\%, 99.05\%; for DenseNet201 97.66\%, 98.05\% for hard and soft respectively. Hence, it's so clear that the soft voting classifier outperforms better than hard voting classifier. So, we select a soft voting classifier in our proposed model for the lung cancer dataset.

\begin{table*}[]
\centering
\begin{tabular}{lllllll}
\hline
Voting Classifer & VGG16 & VGG19 & MobileNet & DenseNet169 & DenseNet201 & Average \\ \hline
Hard & 96.9  & 96.9 & 98.81  & 97.14 & 98.57 & 97.66 \\ 
Soft & 97.14 & 97.14 & 99.05 & 97.86 & 99.05 & 98.05 \\ \hline
\end{tabular}%
\caption{Accuracy performance analysis for ensemble models on lung cancer dataset}
\label{tab:lung_en_table}
\end{table*}

Table \ref{tab:lung_tl_table} shows the performance of different TL models in tabular format and insights that the MobileNet TL model gives the highest average performance among all other TL models where the average accuracy rate for VGG16, VGG19, MobileNet, DenseNet169, and DenseNet201 is 97.02\%, 97.02\%, 98.93\%, 97.5\%, and 98.81\% respectively. Besides in Fig. \ref{fig:lung_en_graph}, we can see that the graphical upper bar chart is pointed to the MobileNet as having the highest accuracy rate among others. So, we select MobileNet in our proposed model for the lung cancer dataset.
\begin{table}[]
\centering
\begin{tabular}{llllll}
\hline
TL & VGG16 & VGG19 & MobileNet & DenseNet169 & DenseNet201 \\ \hline
Hard & 96.9 & 96.9 & 98.81 & 97.14 & 98.57 \\ 
Soft & 97.14 & 97.14 & 99.05 & 97.86 & 99.05 \\ 
Average & 97.02 & 97.02 & 98.93 & 97.5 & 98.81 \\ \hline
\end{tabular}%
\caption{Accuracy performance analysis for TL models on lung cancer dataset}
\label{tab:lung_tl_table}
\end{table}

\begin{figure*}[!htbp]
	\centering
	\subfloat[Performance]{\includegraphics[scale=.550]{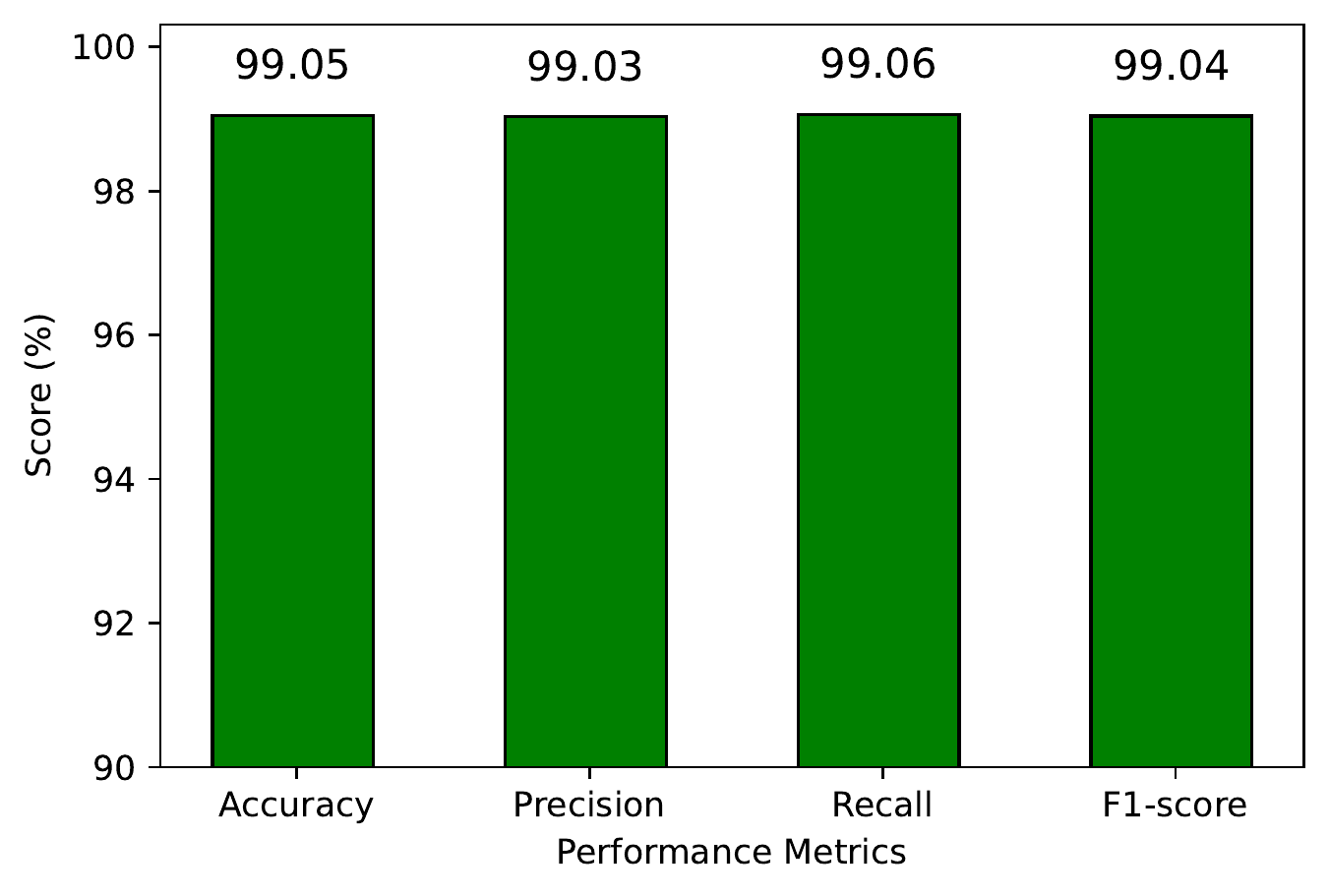}}\hspace{0.1cm}
	\subfloat[Error]{\includegraphics[scale=.550]{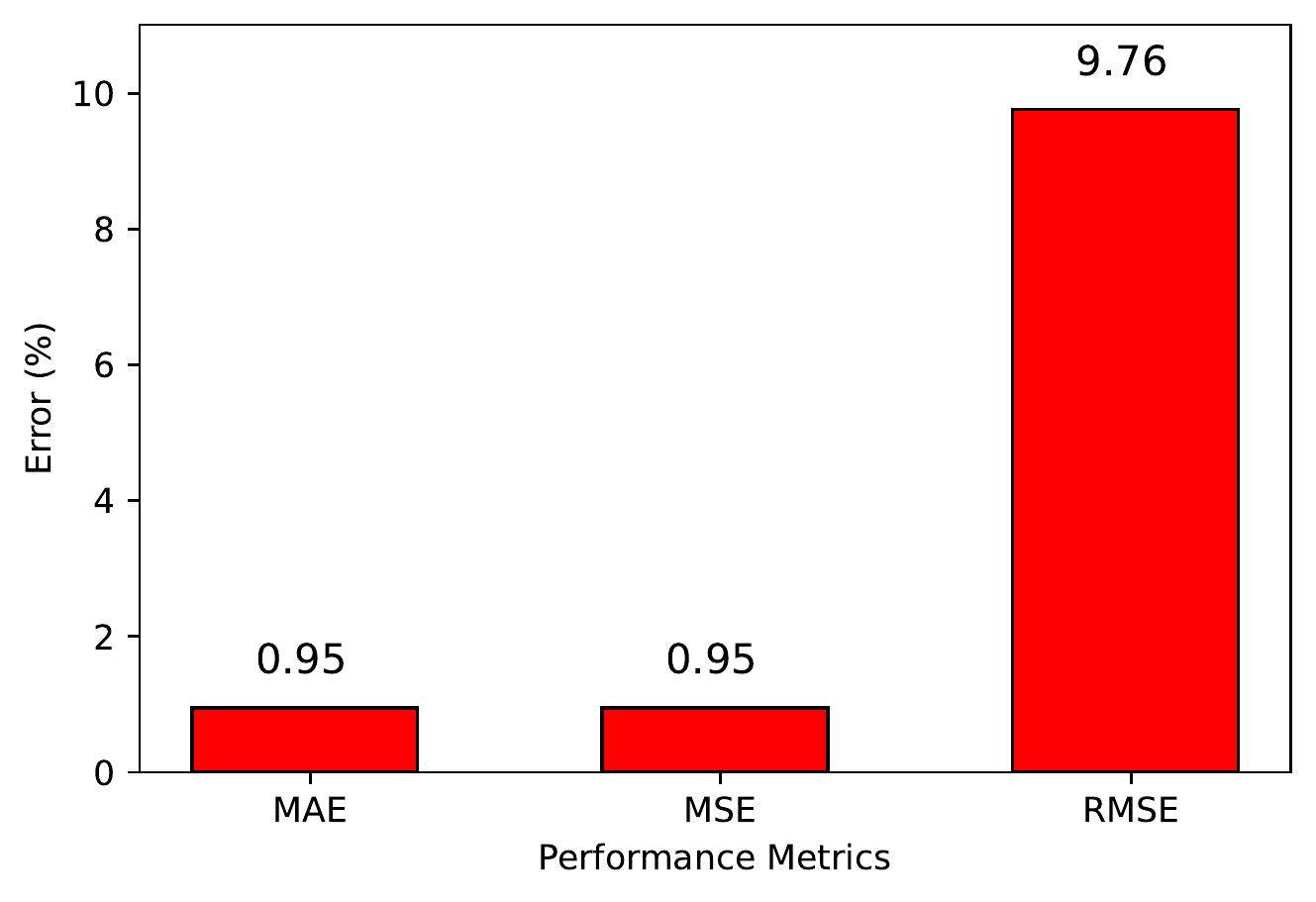}}\hspace{0.1cm}
	\caption{Performance analysis for ensemble soft voting classifier for lung cancer}
	\label{fig:lung_en_pererr_analysis}
\end{figure*}

\begin{figure*}[!htbp]
	\centering
	\subfloat[Confusion Matrix]{\includegraphics[scale=.480]{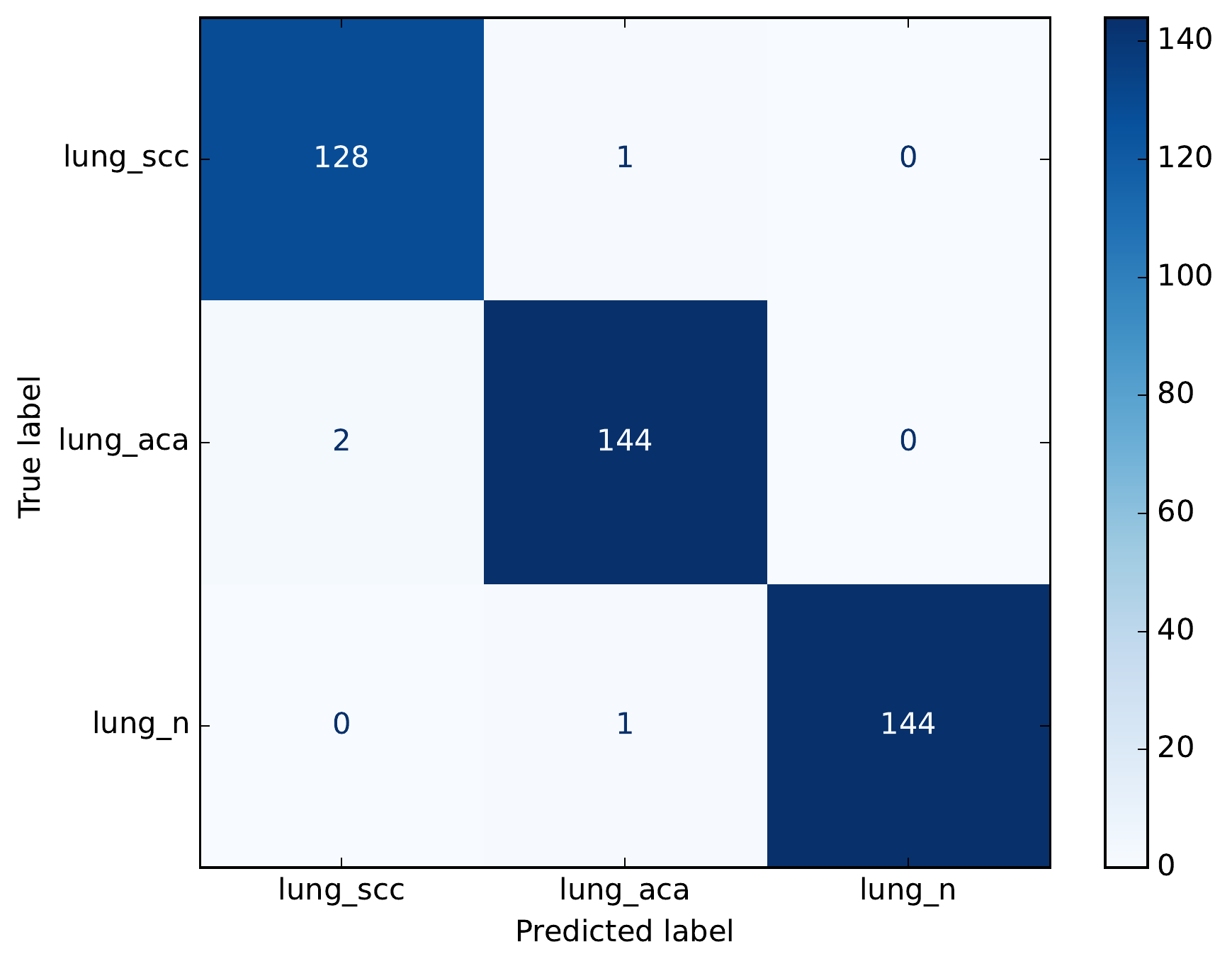}}\hspace{0.1cm} 
	\subfloat[ROC Curve]{\includegraphics[scale=.480]{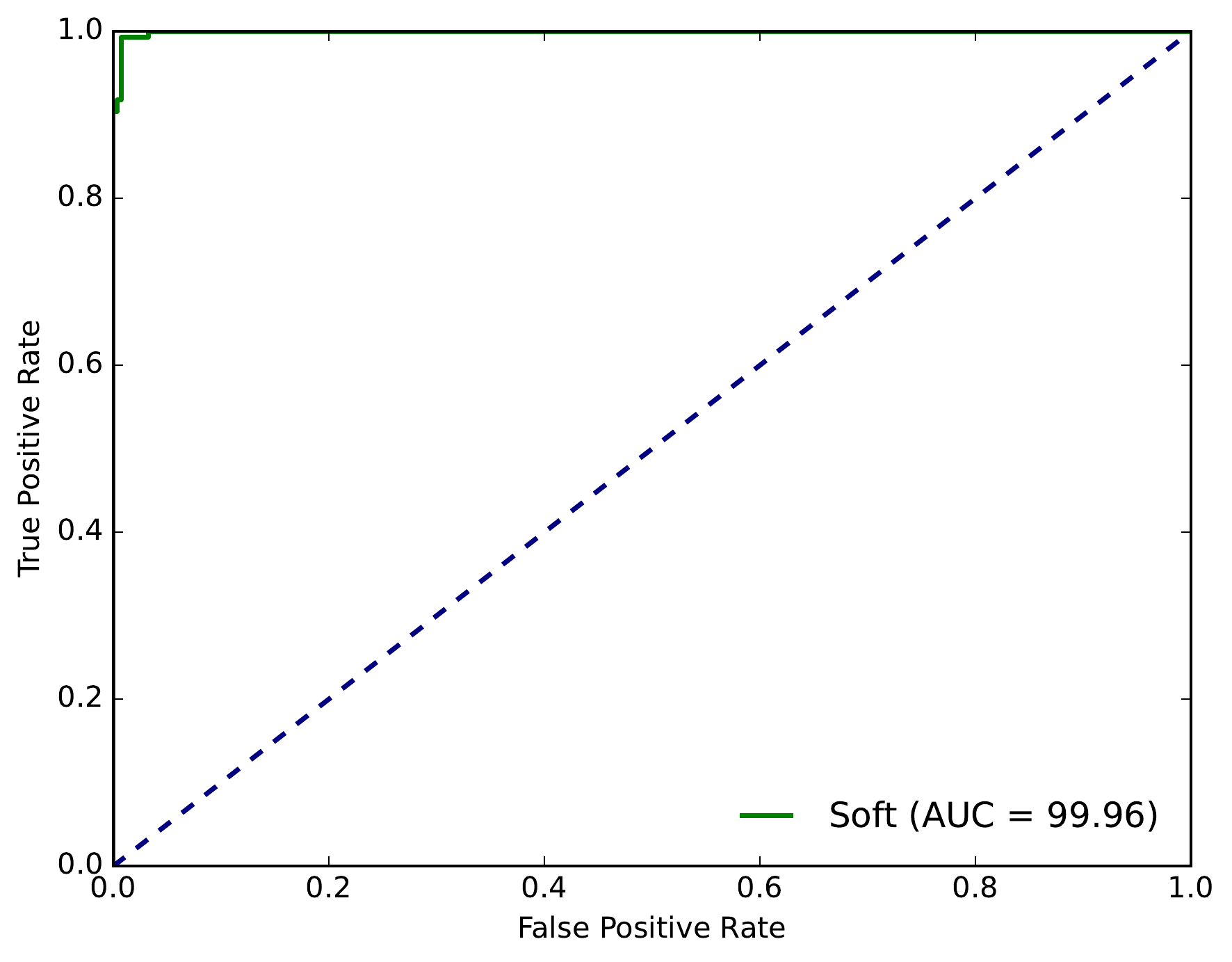}}\hspace{0.1cm}
	\caption{Confusion matrix and ROC Curve for ensemble soft voting classifier for lung cancer}
	\label{fig:lung_en_conroc_analysis}
\end{figure*}

In Fig. \ref{fig:lung_en_pererr_analysis} we can observe the performance of the MobileNet TL model. In performance bar chart, we can visualize that the accuracy, precision, precision, recall, f1-score are 99.05\%, 99.03\%,99.06\%, 99.04\% respectively and in error bar chart, the MAE, MSE, RMSE are 0.95\%,0.95\%, 9.67\% respectively. we can insight that the accuracy rate is much higher and the error rate is much lower. In Fig. \ref{fig:lung_en_conroc_analysis} we can see the confusion matrix and ROC Curve for the MobileNet TL model. In the Confusion Matrix, we can see that the TP rate is 30.48\%, 34.29\%, 34.29\%; TN rate is 68.81\%, 64.76\%, 65.48\%; FP rate is 0.48\%, 0.48\%, 0.0\%; FN rate is 0.24\%, 0.48\%, 0.24\% for lung squamous cell carcinoma, lung adenocarcinomas, and lung benign tissue respectively. Here, we can insight that the true positive and true negative rate is very high, and false positive and false negative are very low. The large number of TP and TN than FP and FN is a very crucial point  for machine learning models to detect lung cancer. In the ROC Curve, we can see that the AUC score is 99.96\% for the soft voting classifier and the curve is very well fitted in the graph as it is close to 1  which is a sign of a better performance model.


\subsection{Performance evaluation of colon cancer}

\begin{figure*}[!htbp]
	\centering
	{\includegraphics[scale=.520]{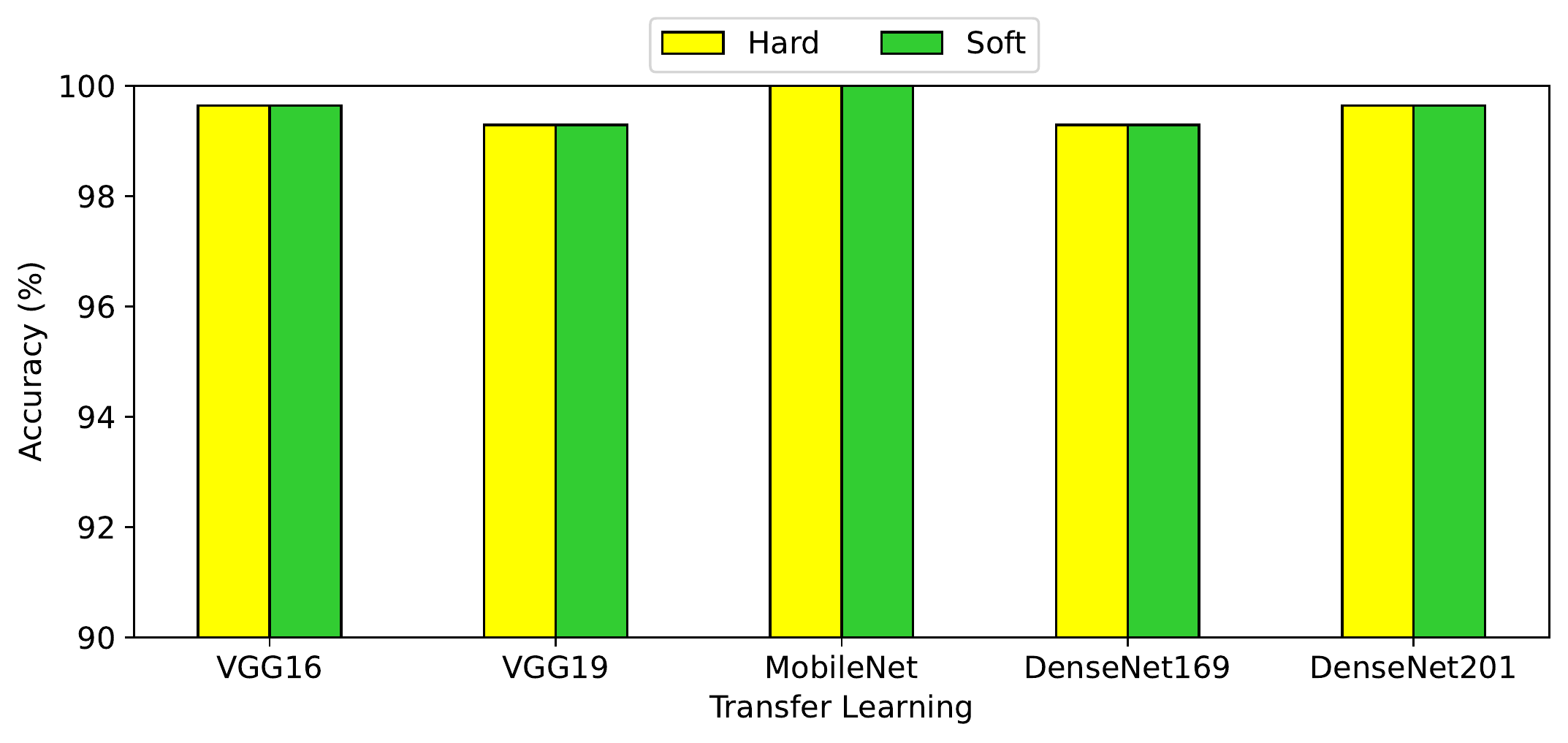}} 
	\caption{Accuracy performance analysis for ensemble models on colon cancer dataset}
	\label{fig:colon_en_graph}
\end{figure*}

Table \ref{tab:colon_en_table} and Fig. \ref{fig:colon_en_graph} illustrate the performance of different TL models based on hard and soft voting in tabular format and visual graphical bar chart respectively. From Table \ref{tab:colon_en_table}, we can see that the average accuracy rate for hard and soft voting classifiers are 99.57\% and 99.57\% respectively. Moreover in Fig. \ref{fig:colon_en_graph} the graphical bar chart, we can see that the graphical upper bar is soft and hard voting classifiers with the same level, where 99.64\%, 99.29\%, 100\%, 99.29\%, 99.64\%, 99.57\% for VGG16, VGG19, MobileNet, DenseNet169, and DenseNet201 respectively for both hard and soft voting classifier. As we know soft voting achieves an accuracy rate based on probabilities in lieu of class labels. So, we select a soft voting classifier in our proposed model for the colon cancer dataset.

\begin{table*}[]
\centering
\begin{tabular}{lllllll}
\hline
Voting Classifier & VGG16 & VGG19 & MobileNet & DenseNet169 & DenseNet201 & Average \\ \hline
Hard & 99.64 & 99.29 & 100 & 99.29 & 99.64 & 99.57 \\ 
Soft & 99.64 & 99.29 & 100 & 99.29 & 99.64 & 99.57 \\ \hline
\end{tabular}%
\caption{Accuracy performance analysis for ensemble models on colon cancer dataset}
\label{tab:colon_en_table}
\end{table*}

Table \ref{tab:colon_tl_table} shows the performance of different TL models in tabular format and insights that the MobileNet TL model gives the highest average performance among all other TL models where the average accuracy rate for VGG16, VGG19, MobileNet, DenseNet169, and DenseNet201 is 99.64\%, 99.29\%, 100\%, 99.29\%, and 99.64\% respectively. Besides in Fig. \ref{fig:colon_en_graph}, we can see that the graphical upper bar chart is pointed to the MobileNet as having the highest accuracy rate among others. So, we select MobileNet in our proposed model for the lung cancer dataset.

\begin{table}[]
\centering
\begin{tabular}{llllll}
\hline
TL & VGG16 & VGG19 & MobileNet & DenseNet169 & DenseNet201 \\ \hline
Hard & 99.64 & 99.29 & 100 & 99.29 & 99.64 \\ 
Soft & 99.64 & 99.29 & 100 & 99.29 & 99.64 \\ 
Average & 99.64 & 99.29 & 100 & 99.29 & 99.64 \\ \hline
\end{tabular}%
\caption{Accuracy performance analysis for TL models on colon cancer dataset}
\label{tab:colon_tl_table}
\end{table}

\begin{figure*}[!htbp]
	\centering
	\subfloat[Performance]{\includegraphics[scale=.550]{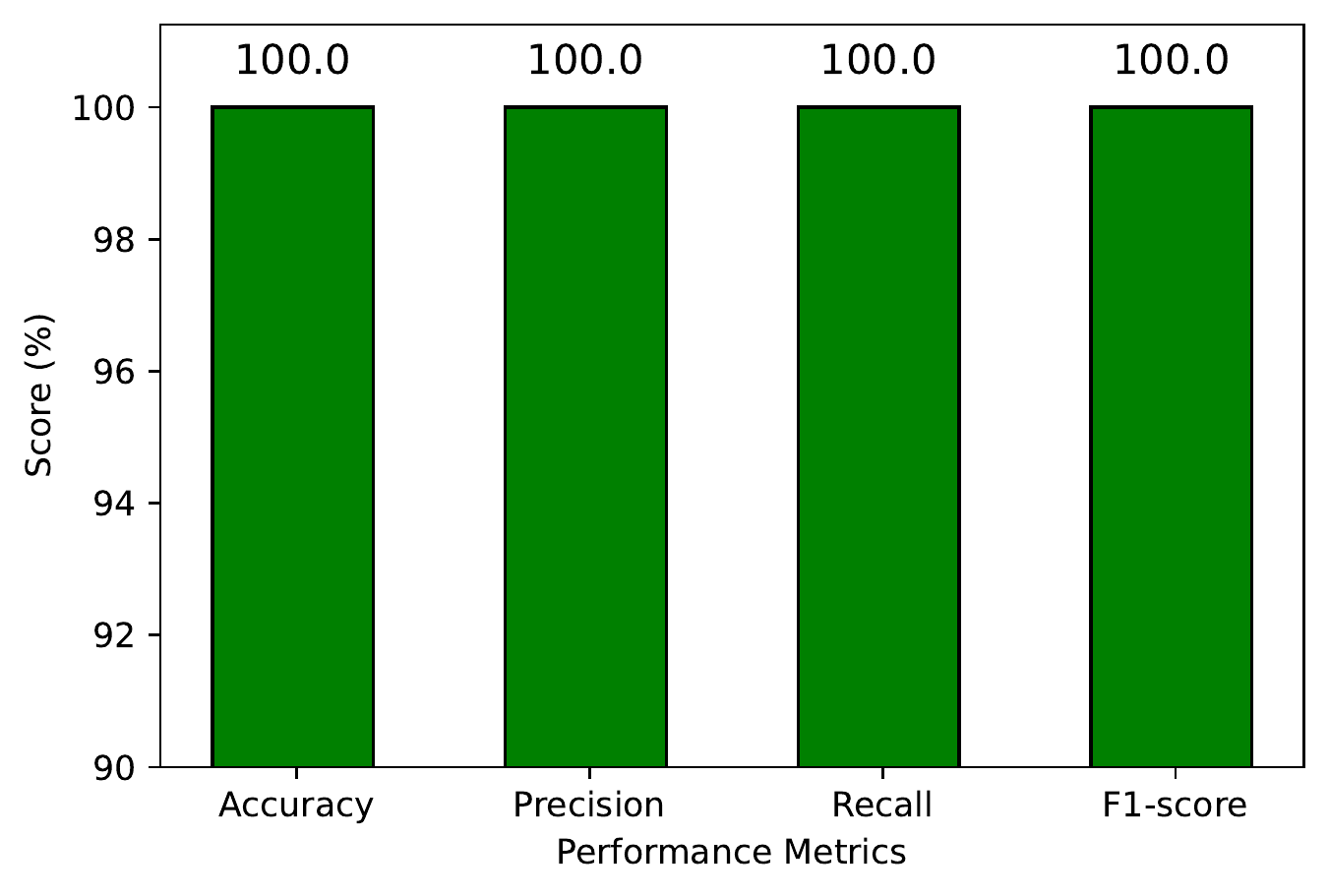}}\hspace{0.1cm}
	\subfloat[Error]{\includegraphics[scale=.550]{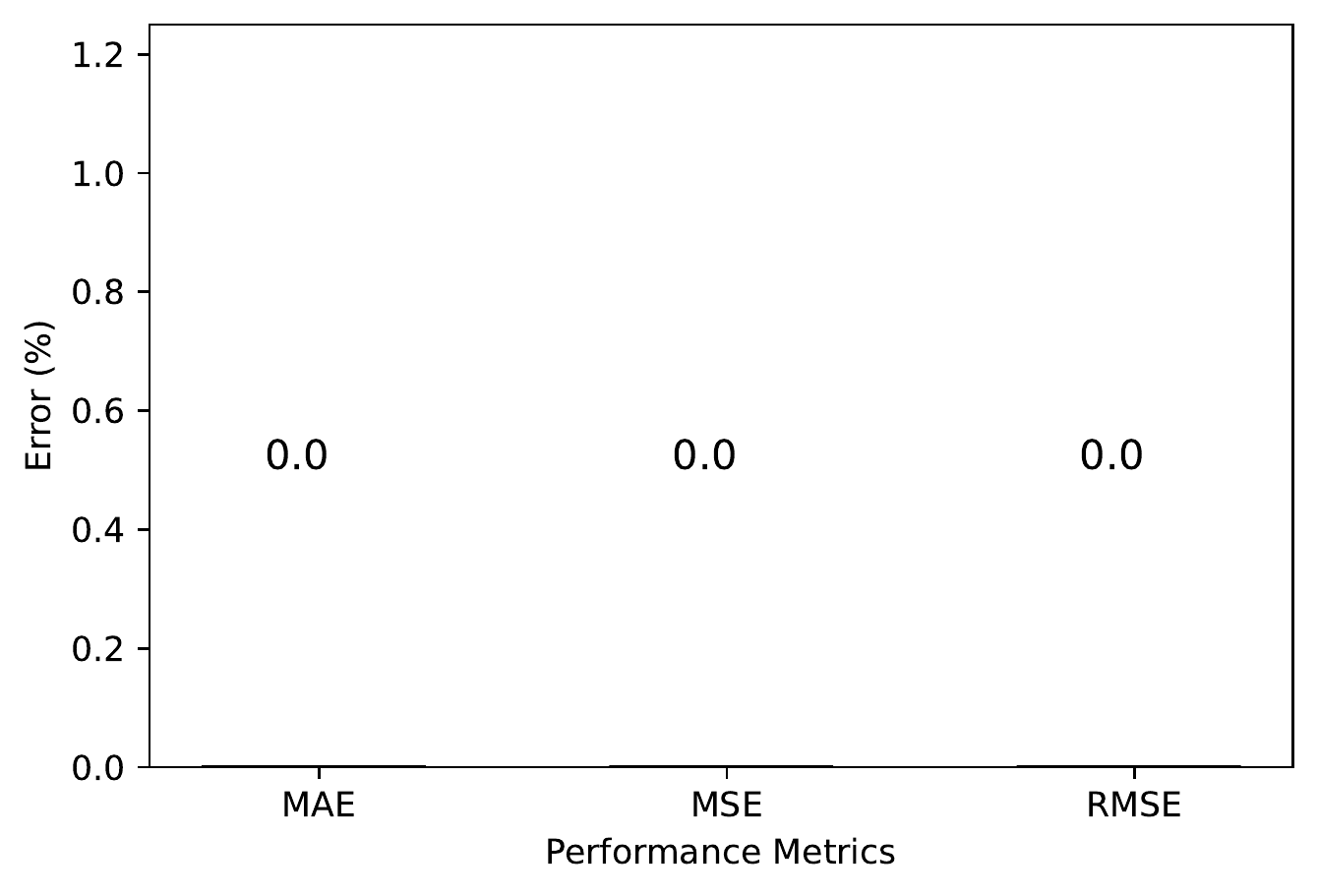}}\hspace{0.1cm}
	\caption{Performance analysis for ensemble soft voting classifier for colon cancer}
	\label{fig:colon_en_pererr_analysis}
\end{figure*}

\begin{figure*}[!htbp]
	\centering
	\subfloat[Confusion Matrix]{\includegraphics[scale=.480]{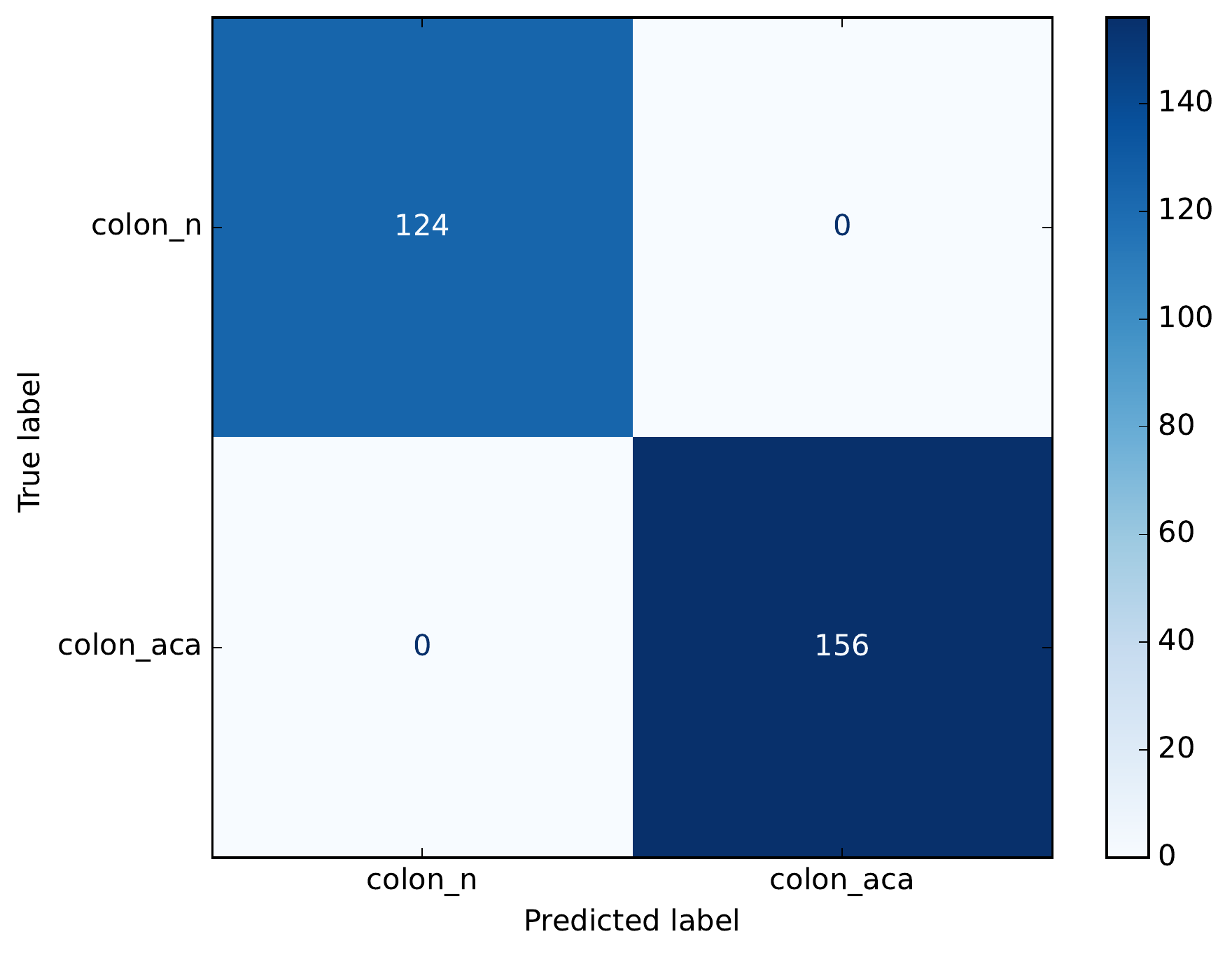}}\hspace{0.1cm}
	\subfloat[ROC Curve]{\includegraphics[scale=.480]{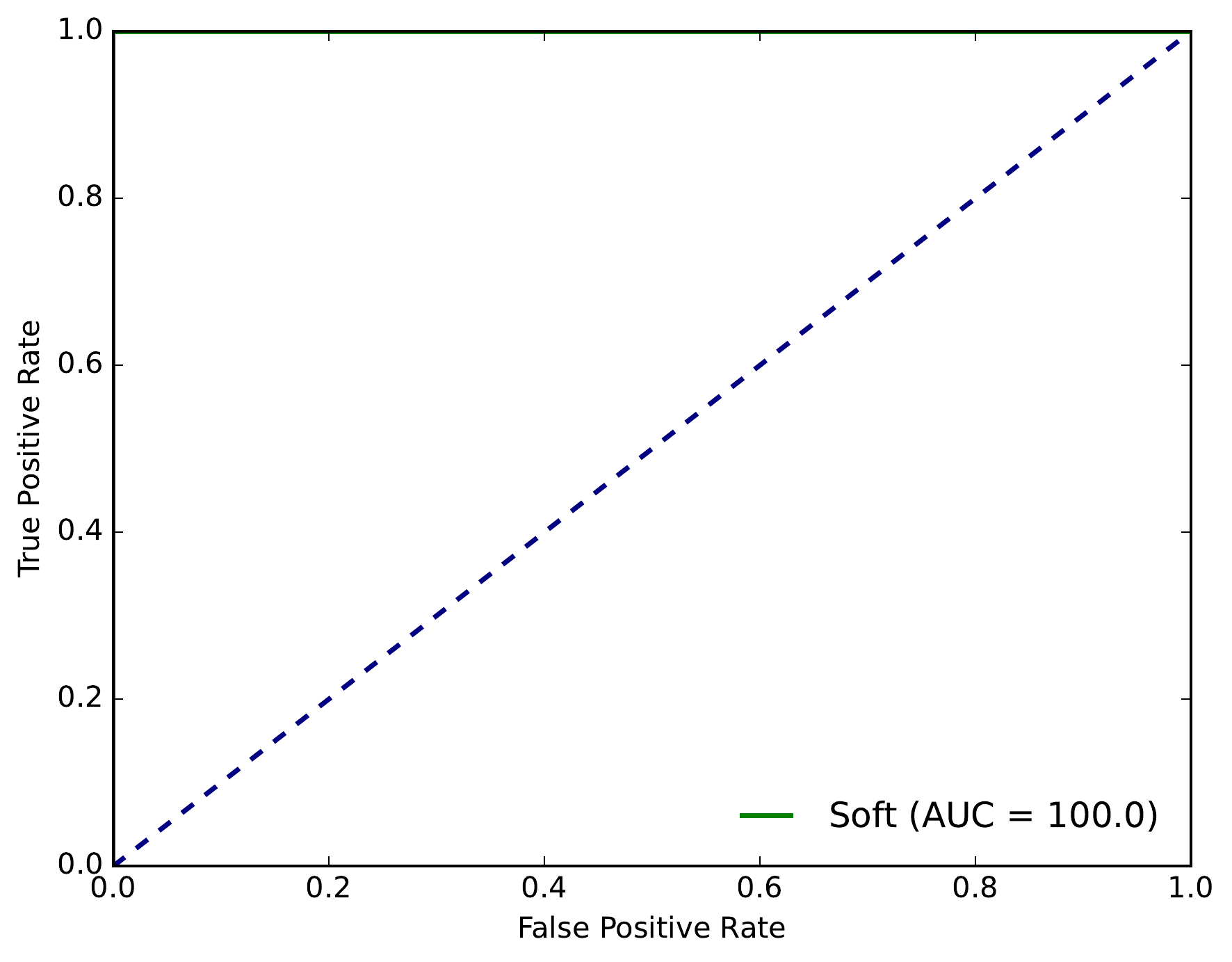}}\hspace{0.1cm}
	\caption{Confusion matrix and ROC Curve for ensemble soft voting classifier for colon cancer}
	\label{fig:colon_en_conroc_analysis}
\end{figure*}

In Fig. \ref{fig:colon_en_pererr_analysis}, we can observe the performance of the MobileNet TL model. In the performance bar chart, we can visualize that the accuracy, precision, precision, recall, and f1-score are 100\% for each and in the error bar chart, the MAE, MSE, and RMSE are 0\% for each. we can insight that the accuracy rate is the highest and the error rate is zero. In Fig. \ref{fig:colon_en_conroc_analysis}, we can see the confusion matrix and ROC Curve for the MobileNet TL model. In the Confusion Matrix, we can see that the TP rate is 44.29\%; TN rate is 55.71\%; FP rate is 0.0\%; FN rate is 0.0\% for colon cancer (colon adenocarcinomas and colon benign tissue) respectively. Here, we can observe that the true positive and true negative rate is very high, and false positive and false negative is very low. The large number of TP and TN than FP and FN is a very crucial point for machine learning models to detect colon cancer. In the ROC Curve, we can see that the AUC score is 100\% for the soft voting classifier and the curve is very well fitted in the graph as it is 1 which is a sign of a perfect performance model.

The accuracy rate of colon cancer is achieved at 100\% which is a very promising result ever in colon cancer detection. With the help of proper pre-processing, feature extraction and efficient ensemble soft voting classifier give these good results. Among all transfer learning models, the MobileNet model gives better feature extraction than others so the accuracy label reached the highest ever. As we already know that ensemble learning is a multiple classifiers system \citep{gudivada2016cognitive}. It's better to get a prediction from multiple models than a single model, which makes a model more robust and efficient. Hence, we have used the ensemble method to get a pleasant model with good performance results.

\subsection{Performance evaluation of cancer (lung and colon)}

\begin{figure*}[!htbp]
	\centering
	{\includegraphics[scale=.520]{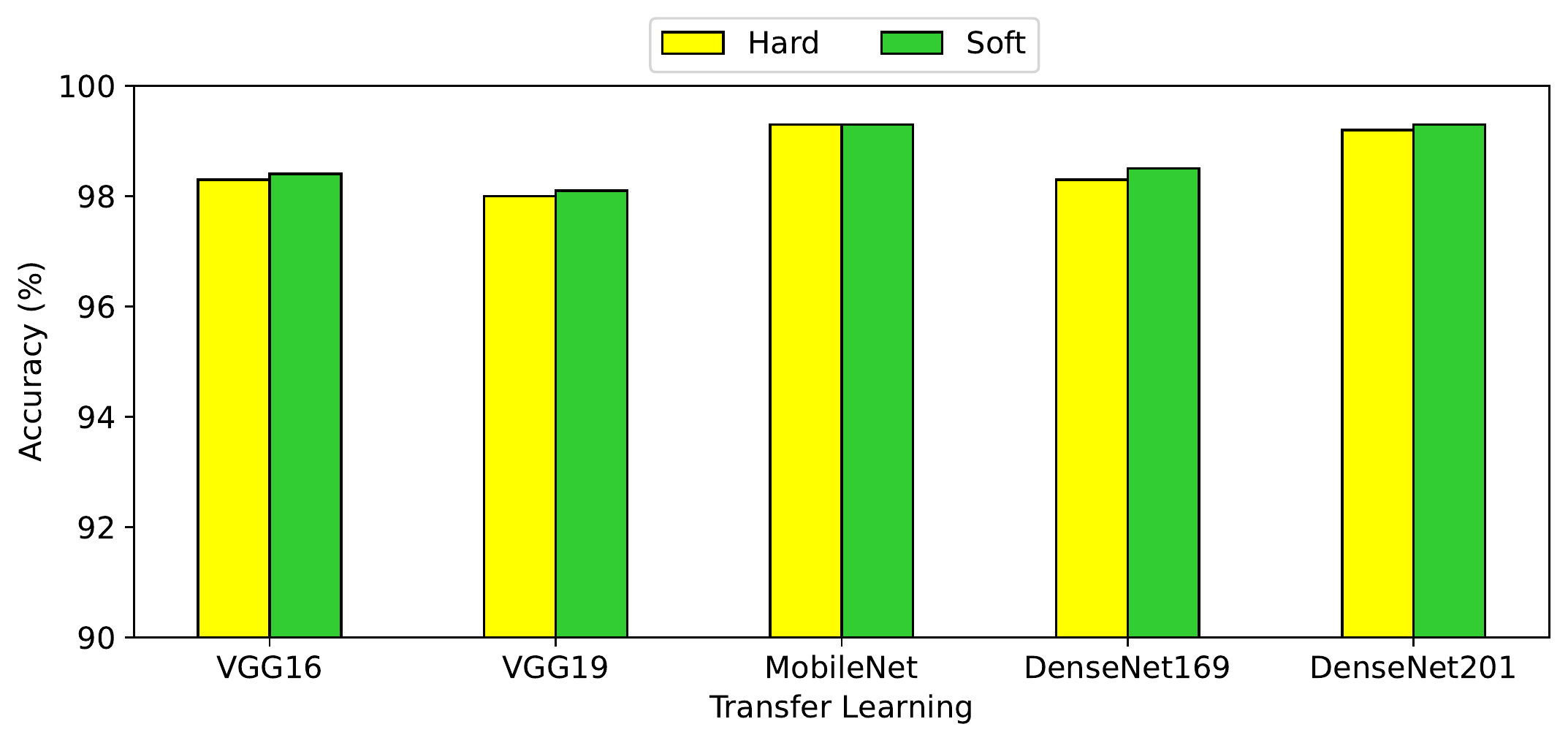}}
	\caption{Accuracy performance analysis for ensemble models on  (lung \& cancer) cancer dataset}
	\label{fig:cancer_en_graph}
\end{figure*}

Table \ref{tab:cancer_en_table} and Fig. \ref{fig:cancer_en_graph} illustrate the performance of different TL models based on hard and soft voting in tabular format and visual graphical bar chart respectively. From Table \ref{tab:cancer_en_table}, we can see that the average accuracy rate for hard and soft voting classifiers is 98.62\% and 98.72\% respectively. Moreover in Fig. \ref{fig:cancer_en_graph} for the graphical bar chart, we can see that the graphical upper bar is soft and lower is hard voting classifier, where for VGG1698.3\%, 98.4\%; for VGG19 98.0\%, 98.1\%; for MobileNet 99.30, 99.30\%; for DenseNet169 99.3\%,98.5\%; for DenseNet201 99.2\%, 99.3\% for hard and soft respectively. Hence, it's so clear that the soft voting classifier outperforms better than hard voting classifier. So, we select a soft voting classifier in our proposed model for the (lung and colon) cancer dataset.

\begin{table*}[]
\centering
\begin{tabular}{lllllll}
\hline
Voting Classifer & VGG16 & VGG19 & MobileNet & DenseNet169 & DenseNet201 & Average \\ \hline
Hard & 98.3 & 98.0 & 99.3 & 98.3 & 99.2 & 98.62 \\ 
Soft & 98.4 & 98.1 & 99.3 & 98.5 & 99.3 & 98.72 \\ \hline
\end{tabular}%
\caption{Accuracy performance analysis for ensemble models on (lung and colon) cancer dataset}
\label{tab:cancer_en_table}
\end{table*}

Table \ref{tab:cancer_tl_table} shows the performance of different TL models in tabular format and insights that the MobileNet TL model gives the highest average performance among all other TL models where the average accuracy rate for VGG16, VGG19, MobileNet, DenseNet169, and DenseNet201 is 98.35\%, 98.05\%, 99.30\%, 98.40\%, and 99.25\% respectively. Besides in Fig. \ref{fig:cancer_en_graph}, we can see that the graphical upper bar chart is pointed to the MobileNet as having the highest accuracy rate among others. So, we select MobileNet in our proposed model for the (lung and colon) cancer dataset.

\begin{table}[]
\centering
\begin{tabular}{llllll}
\hline
TL & VGG16 & VGG19 & MobileNet & DenseNet169 & DenseNet201 \\ \hline
Hard & 98.3 & 98.0 & 99.3 & 98.3 & 99.2 \\ 
Soft & 98.4 & 98.1 & 99.3 & 98.5 & 99.3 \\ 
Average & 98.35 & 98.05 & 99.3 & 98.4 & 99.25 \\ \hline
\end{tabular}%
\caption{Accuracy performance analysis for TL models on cancer (lung and colon) dataset}
\label{tab:cancer_tl_table}
\end{table}

\begin{figure*}[!htbp]
	\centering
	\subfloat[Performance]{\includegraphics[scale=.550]{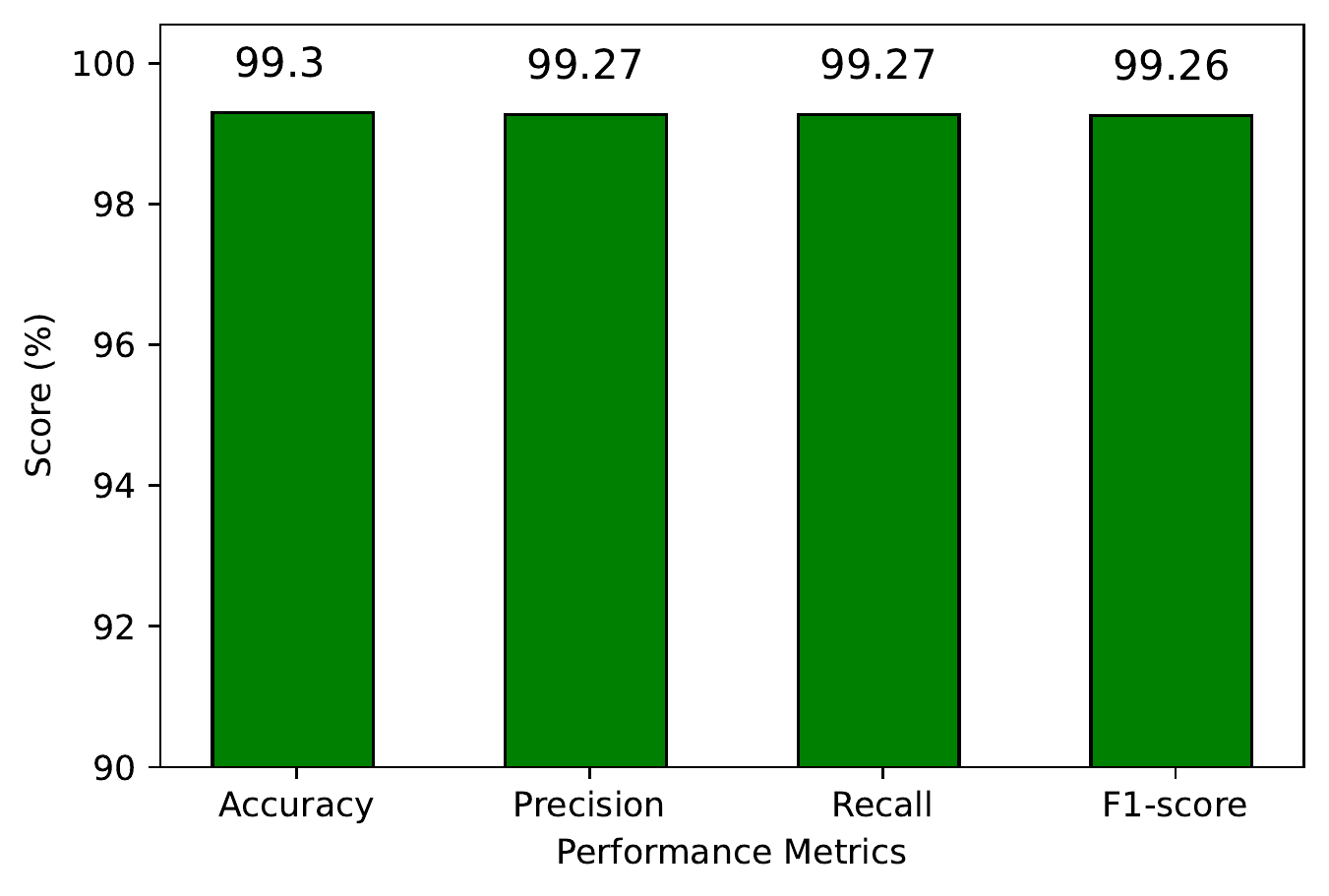}}\hspace{0.1cm}
	\subfloat[Error]{\includegraphics[scale=.550]{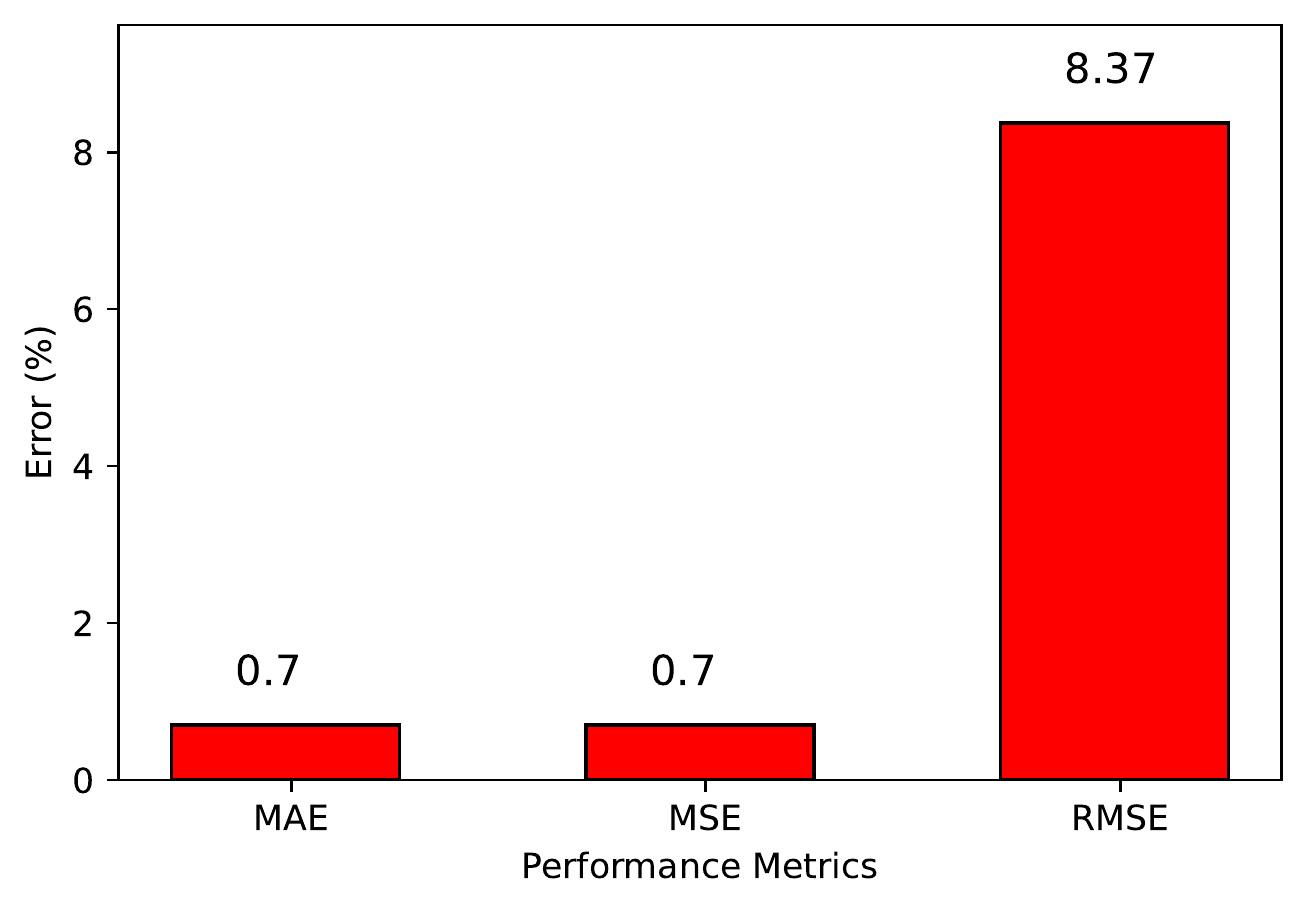}}\hspace{0.1cm}
	\caption{Performance analysis for ensemble soft voting classifier for (lung and colon) cancer}
	\label{fig:cancer_en_pererr_analysis}
\end{figure*}

\begin{figure}[!htbp]
	\centering
	\subfloat[Confusion Matrix]{\includegraphics[scale=.440]{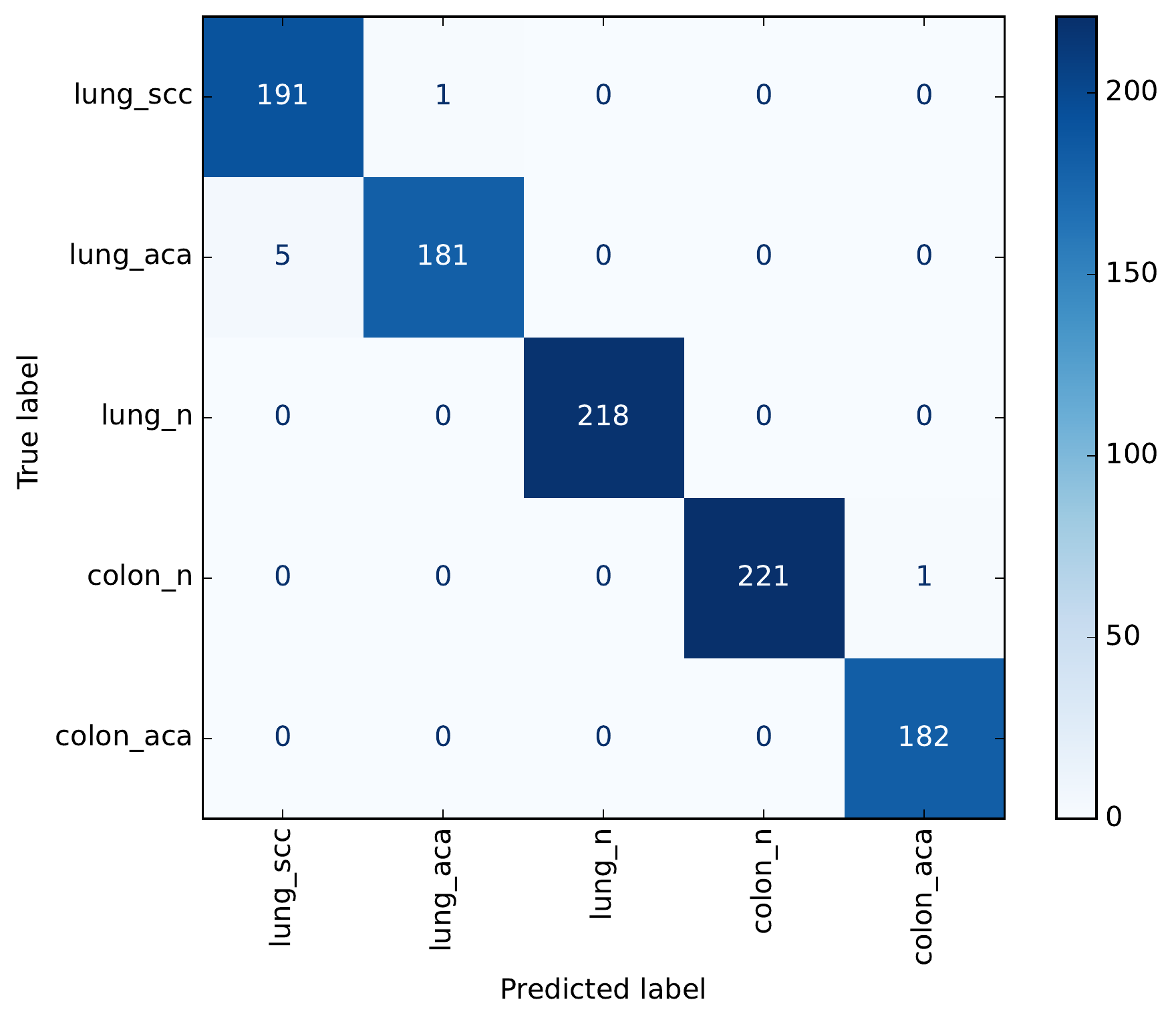}}\hspace{0.1cm}
	\subfloat[ROC Curve]{\includegraphics[scale=.480]{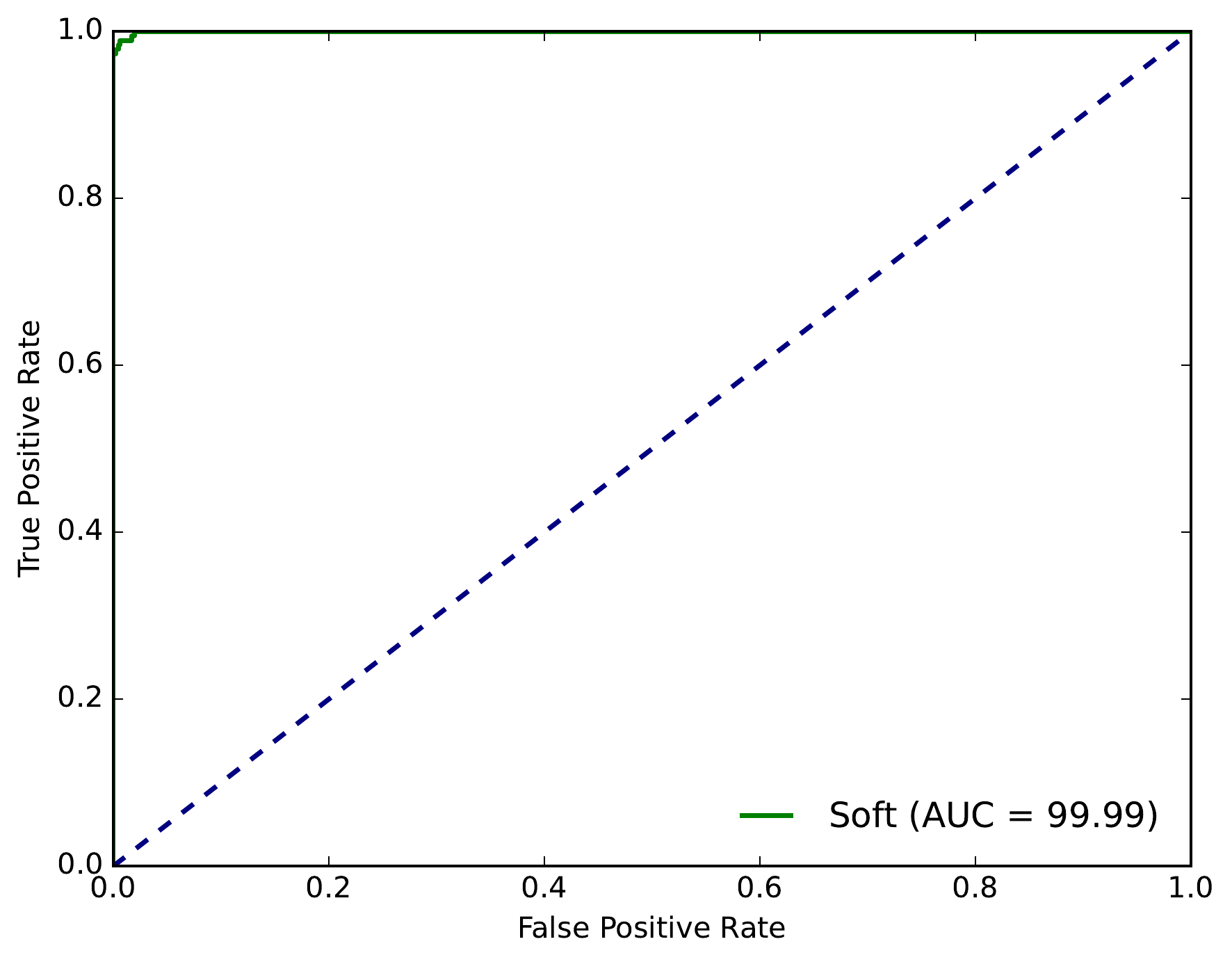}}\hspace{0.1cm}
	\caption{Confusion matrix and ROC Curve for ensemble soft voting classifier for (lung and colon) cancer}
	\label{fig:cancer_en_conroc_analysis}
\end{figure}

In Fig. \ref{fig:cancer_en_pererr_analysis}, we can observe the performance of the MobileNet TL model. In performance bar chart, we can visualize that the accuracy, precision, precision, recall, f1-score are 99.30\%, 99.27\%,99.27\%, 99.26\% respectively and in error bar chart, the MAE, MSE, RMSE are 0.7\%,0.7\%, 8.37\% respectively. we can insight that the accuracy rate is much higher and the error rate is much lower. In Fig. \ref{fig:cancer_en_conroc_analysis}, we can see the confusion matrix and ROC Curve for the MobileNet TL model. In the Confusion Matrix, we can see that the TP rate is 19.1\%, 18.1\%, 21.8\%, 22.1\%, 18.2\%; TN rate is 80.3\%, 81.3\%, 78.2\%, 77.8\%, 81.7\%; FP rate is 0.5\%, 0.1\%, 0.0\%, 0.0\%, 0.1\%; FN rate is 0.1\%, 0.5\%, 0.0\%, 0.1\%, 0.0\% for lung squamous cell carcinoma, lung adenocarcinomas, lung benign, colon benign, and colon adenocarcinomas respectively. Here, we can observe that the true positive and true negative rate is very high, and false positive and false negative is very low. The large number of TP and TN than FP and FN is a very crucial point  for machine learning models to detect lung cancer. In the ROC Curve, we can see that the AUC score is 99.99\% for the soft voting classifier and the curve is very well fitted in the graph as it is close to 1 which is a sign of a better performance model.

\subsection{Discussion}

In our proposed model, we have evaluated our model using deep feature extraction and ensemble learning approach based on high-performance filtering and locating the best TL models for feature extraction as well as ensemble models providing a better accuracy rate.   The proposed method is compared with the methods in other papers and significantly achieves the highest accuracy than others. The comparison study of our proposed model with others is illustrated in tabular format in Table \ref{tab:comparision_analysis}. In comparison, the dataset that we have utilized and compared with others is the same dataset (HPI LC25000) and the number of images employing the methods is slightly different. The similar datasets with a tiny variation of datasets make the comparison as a fair comparison to estimate the performance.
 
In our experiments, we achieve the highest performance accuracy rate: lung dataset 99.05\%; colon dataset 100\%; (lung and colon) dataset 99.30\% which outperforms the others.
Hence, Among all of the preceding in both lung and colon cancer detection, our proposed hybrid model significantly achieves the highest accuracy.


\begin{table}[]
\centering
\resizebox{\textwidth}{!}{%
\begin{tabular}{lllll}
\hline
SI. No. & Author & Approach & Dataset & Accuracy (In \%) \\ \hline
1 & \cite{hatuwal2020lung} & CNN & \multirow{4}{*}{HPI Lung \citep{borkowski2019lung}} & 97.20 \\
2 & \cite{mangal2020convolution} & SCNN &  & 97.89 \\
3 & \cite{shandilya2022analysis} & CAD &  & 98.67 \\
4 & Our proposed model & Hybrid Model &  & \textbf{99.05} \\
5 & \cite{tasnim2021deep} & \begin{tabular}[c]{@{}l@{}}CNN (MaxPool)\\ CNN (AvgPool)\\ MobileNetV2\end{tabular} & \multirow{6}{*}{HPI Colon \citep{borkowski2019lung}} & \begin{tabular}[c]{@{}l@{}}97.49 \\ 95.48 \\ 99.67\end{tabular} \\
6 & \cite{liang2020identification} & MFF-CNN &  & 96 \\
7 & \cite{mangal2020convolution} & SCNN &  & 96.61 \\
8 & \cite{qasim2020convolutional} & CNN &  & 99.6 \\
9 & \cite{yildirim2022classification} & MA ColonNET &  & 99.75 \\
10 & Our proposed model & Hybrid Model &  & \textbf{100} \\ 
11 & \cite{masud2021machine} & 2D FW+ CNN & \multirow{4}{*}{HPI Lung and Colon \citep{borkowski2019lung}} & 96.33 \\
12 & \cite{sikdersupervised} & CNN &  & 93 \\
13 & \cite{adu2021dhs} & DHS-CapsNet &  & 99.23 \\
14 & Our proposed model & Hybrid Model &  & \textbf{99.30} \\ \hline
\end{tabular}%
}
\caption{Comparison analysis of lung and colon cancer}
\label{tab:comparision_analysis}
\end{table}

\subsection{Speed and complexity analysis}
The prediction speed is the time needs to predict cancer and the resource usage cost is simply the complexity of the ML model. The time and space complexity, time requires to execute an operation and the space requires to store data, are very important to determine the efficiency of any algorithm \citep{zhu2022eco} \citep{abdelsamea2019cascade}. In the following section, we have mentioned the prediction speed and complexity analysis.
The diagnosis of lung and colon cancer has an overall computational complexity of O(n²). In our method, we take the estimators from the HPF process that requires O(n²) and build an ensemble learning method that consumes O(n) time to complete, so the overall complexity is O(n²) + O(n) $\sim$ O(n²). The other ML models such as RF, SVM, LR, MLP, XGB, and LGB take O(tlogn), O(n²), O(m), O(n²), O(td), and O(tl) respectively, where n is the number of data points, m is the number of features, t denotes the number of trees, l is the level of the trees, and d is the depth of the trees. On the other hand, diagnosing lung and colon cancer has an overall space complexity of O(n²). In our method, we take the estimators from the HPF process that requires O(n²) and build an ensemble learning method that consumes O(n) time to complete, so the overall complexity is O(n²) + O(n) $\sim$ O(n²). The other ML models such as RF, SVM, LR, MLP, XGB, and LGB take O(et), O(km), O(m), O(n²), O(et+g), and O(et+l) respectively, where n is the number of data points, m is the number of features, t denotes the number of trees, e is the number of nodes, l is the level of the trees, and g is the output values for each leaf in decision trees. The analysis of time and space complexity is shown in Table \ref{tab:time_space}.
\begin{table}[!h]
\centering
\begin{tabular}{clcccccccccc}
\hline
\multicolumn{2}{c}{Method} & RF & SVM & LR & MLP & XGB & LGB & HPF& EL& Proposed Model \\ \hline
\multicolumn{2}{c}{Time Complexity} & O(tlogn) & O(n²) & O(m) & O(n²) & O(td) & O(tl)& O(n²) & O(n)& O(n²) \\
\multicolumn{2}{c}{Space Complexity} & O(et) & O(km) & O(m) & O(n²) & O(et+g) & O(et+g)& O(n²) & O(n)& O(n²) \\ \hline
\end{tabular}
\caption{The time and space complexity of the ML Models}
\label{tab:time_space}
\end{table}
Furthermore, we estimated the prediction time (sec) on the system having 4 cores CPU, 16GB RAM, and 1 GPU employed by our proposed model while working with the 420 (lung), 280 (colon), and 1000 (lung and colon) testing images. For lung, colon, and (lung and colon) cancer, our approach encloses the prediction times of 2s, 0s, and 14s, respectively. The prediction time grows as the number of testing images grows. The analysis of prediction time is shown in Table \ref{tab:pred_time}.
\begin{table}[!h]
\centering
\resizebox{\textwidth}{!}{%
\begin{tabular}{ccccccccccc}
\hline
\multicolumn{2}{c}{Method} & RF & SVM & LR & MLP & XGB & LGB & Proposed Model \\ \hline
\multirow{3}{*}{Prediction Time (sec)} 
&Lung Dataset & 0 & 2 & 0 & 0 & 0 & 0  & 2 \\
&Colon Dataset & 0 & 0 & 0 & 0 & 0 & 0 & 0 \\
&Lung and Colon Dataset & 0 & 14 & 0 & 0 & 0 & 0 & 14\\ \hline
\end{tabular}
}%
\caption{The prediction time (sec) of the ML Models}
\label{tab:pred_time}
\end{table}

\section{Conclusion}
\label{sec:conclusion}
This paper presented a hybrid model for lung and colon cancer detection that integrates pre-processing, k-fold (10 fold) cross-validation, feature extraction using several TL models, devoting ML algorithms, picking the best algorithms using HPF and suit into ensemble learning models. After that select the best ensemble voting classifier model and most suitable feature extraction TL model. Moreover, we employed five TL models in feature extraction such as VGG16, VGG19, MobileNet, DenseNet169, and DenseNet201.  At the same time, for evaluating the performance using ML algorithms, we used six popular algorithms namely RF, SVM, LR, MLP, XGB, and LGB. This extensive development has been evaluated by a set of performance metrics such as accuracy, recall, precision, f1-score, ROC Curve, MAE, MSE, and RMSE. After considering our experiments,  we choose the MobileNet model for Feature extraction among all TL models, and an ensemble soft voting classifier from two ensemble models to evaluate our performance. Our experiment uses the LC25000 lung and colon histopathological image datasets and finds that our hybrid model is best suited for diagnosing lung and colon cancer. In terms of performance metrics, our suggested model performed admirably and outperformed other techniques. The accuracy rate for lung cancer detection is 99.05\%; for colon cancer, it is 100\%, and for lung and colon cancer, it is 99.30\%. We believe our developed model could be applicable in clinics for the automated diagnosis of lung and colon cancers. Even though the architecture delivers greater accuracy, further sophisticated work on image pre-processing and the proposed model would make the task more adequate. A good feature extraction leads to producing better performance results if the image is clear and sharp to extract features is an inadequacy in this paper. In the future, we will go to explore the newly available lung and colon cancer dataset with efficient preprocessing to build a deep learning approach to enhance the performance to detect lung and colon cancer. 

 
\bibliographystyle{apa}
\biboptions{authoryear}

\bibliography{bibs}
\end{document}